\documentclass[lettersize,journal]{IEEEtran}
\usepackage{amsmath,amsfonts}
\usepackage{algorithmic}
\usepackage{algorithm}
\usepackage{array}
\usepackage[caption=false,font=normalsize,labelfont=sf,textfont=sf]{subfig}
\usepackage{textcomp}
\usepackage{stfloats}
\usepackage{url}
\usepackage{verbatim}
\usepackage{graphicx}
\usepackage{cite}
\usepackage{booktabs,multirow}
\usepackage{amssymb}
\usepackage{threeparttable} % To use footnotes within tables
\usepackage{xcolor}

\usepackage{graphicx}  
\usepackage{float}

\newtheorem{definition}{Definition}

\newtheorem{theorem}{Theorem}
\newtheorem{proof}{Proof}

\begin{document}

\title{Alleviating Non-identifiability: a High-fidelity Calibration Objective for Financial Market Simulation with Multivariate Time Series Data}

% \author{IEEE Publication Technology,~\IEEEmembership{Staff, IEEE}
% \thanks{}(Corresponding author: Peng Yang)}

\author{Chenkai Wang,
        Junji Ren,
        and Peng Yang,~\IEEEmembership{Senior Member, IEEE}% <-this % stops a space
\thanks{This paper was produced by the IEEE Publication Technology Group. They are in Piscataway, NJ. Manuscript received XXX, XXX. This work was supported by the National Natural Science Foundation of China (Grants 62250710682, 62272210, and 62331014). (Corresponding author: Peng Yang).}
\thanks{Chenkai Wang is with Department of Statistics and Data Science, Southern University of Science and Technology, Shenzhen 518055, China (email: wangck2022@mail.sustech.edu.cn).}% <-this % stops a space
\thanks{Junji Ren and Peng Yang are with the Guangdong Provincial Key Laboratory of Brain-Inspired Intelligent Computation and the Department of Computer Science and Engineering, Southern University of Science and Technology, Shenzhen 518055, China (emails: 12432695@mail.sustech.edu.cn; yangp@sustech.edu.cn). Peng Yang is also with the Department of Statistics and Data Science at the same university.}
}

% \markboth{Journal of IEEE Transactions on Computational Social Systems,~Vol.~14, No.~8, August~2021}%
\markboth{IEEE Transactions on Computational Social Systems}%
{Shell \MakeLowercase{\textit{et al.}}: A Sample Article Using IEEEtran.cls for IEEE Journals}


\maketitle

\begin{abstract}
The non-identifiability issue has been frequently reported in social simulation works, where different parameters of an agent-based simulation model yield indistinguishable simulated time series data under certain discrepancy metrics. This issue largely undermines the simulation fidelity yet lacks dedicated investigations. This paper theoretically demonstrates that incorporating multiple time series data features during the model calibration phase can exponentially alleviate non-identifiability as the number of features increases. To implement this theoretical finding, a maximization-based aggregation function is proposed based on existing discrepancy metrics to form a new calibration objective function. For verification, the task of calibrating the Financial Market Simulation (FMS), a typical yet complex social simulation, is considered. Empirical studies confirm the significant improvements in alleviating the non-identifiability of calibration tasks. Furthermore, as a model-agnostic method, it achieves much higher simulation fidelity of the chosen FMS model on both synthetic and real market data. Moreover, it is both theoretically and empirically analyzed that as long as the features are selected and not linearly correlated, they can contribute to alleviation, which demonstrates the robustness of the proposed objective. Hence, this work is expected to provide not only a rigorous understanding of non-identifiability in social simulation but also an off-the-shelf high-fidelity calibration objective function for FMS.
\end{abstract}


\section{Introduction}\label{section:intro}
Social simulation plays a pivotal role in understanding complex social systems in a way of revealing their endogenous complexity \cite{squazzoni2014social}. By properly modeling the endogenous components of the social system as multiple interacting agents, i.e., using the Agent-based Models (ABMs) \cite{farmer2009economy}, various what-if tests can be conducted by intervening the agents of interest and analyzing the evolution process of the simulators \cite{fabretti2013problem,weinstein2009meltdown}. These what-if tests involve simulating hypothetical scenarios to assess how changes in agent behaviors or interactions may influence the overall system dynamics.

The success of a social simulation largely lies in the model's fidelity. Note that the what-if tests are often counterfactual, i.e., they do not happen in real social systems, and thus, there are no grounds for validation. To what extent should we believe the results of what-if tests? Besides, due to the uncertainty of human activities, it is often impractical to derive mathematically trustworthy social simulation models. Usually, this issue is addressed in a data-driven manner via model calibration \cite{kim2021automatic}. 

Specifically, a simulation model $M(\omega)$ is essentially a data-generating process that can produce sequences of arbitrary length, i.e., $M(\omega)=\mathbf{X}_{T^{\prime}}$, where $T^{\prime}=[1,...,\tau]$ and $\tau \in \mathbb{N}^+$. If the simulated data $\mathbf{X}_{T}$ resembles the observed data $\hat{\mathbf{X}}_T$ of the targeted real system within any discrete time interval of interest $T=[t_s,t_s+1,..., t_e]$, we consider that $M(\omega)$ has learned the underlying data-generating probabilistic distribution of the real system for generating $\hat{\mathbf{X}}_T$. Therefore, when using $M(\omega)$ for what-if tests, by keeping the non-intervened agents of $M(\omega)$ unchanged, the simulation results of how the intervened agent influences the system are considered trustworthy. To obtain such $M(\omega)$, the discrepancy between $\mathbf{X}_T$ and $\hat{\mathbf{X}}_T$ is measured by various metrics $D$ \cite{platt2020comparison}. The smaller $D$ is, the closer $M(\omega)$ approximates the data-generating distribution of the real social system. By fixing the ABM-based model structure $M$, calibrating the parameter $\omega$ to fit a given observed data sequence $\hat{\mathbf{X}}_T$ is often modeled as an optimization problem \cite{platt2020comparison, yang2025towards}:
\begin{equation}\label{eq:objective function}
\min _{\omega \in \Omega^{D}} D\left(\hat{\mathbf{X}}_T, M\left(\omega\right)=\mathbf{X}_T\right).
\end{equation}

In practice, the calibration methods often suffer from the so-called non-identifiability issue \cite{goosen2021calibrating,bai2022efficient}. That is, there are multiple parameters in the parameter space $\Omega^{D}$ share the same discrepancy value to the given $\hat{\mathbf{X}}_T$. However, those parameters highly likely lead to different distributions of data generation due to the nonlinear model structure $M(\omega)$. In other words, the “ground-truth” parameter of deciding the “ground-truth” data generating distribution cannot be effectively identified from the other parameters in those sets. Consequently, the what-if tests on the resultant model with a randomly picked non-identifiable parameter from there are less trustworthy.

\begin{figure}[t]
  \centering
  \includegraphics[width=0.95\linewidth]{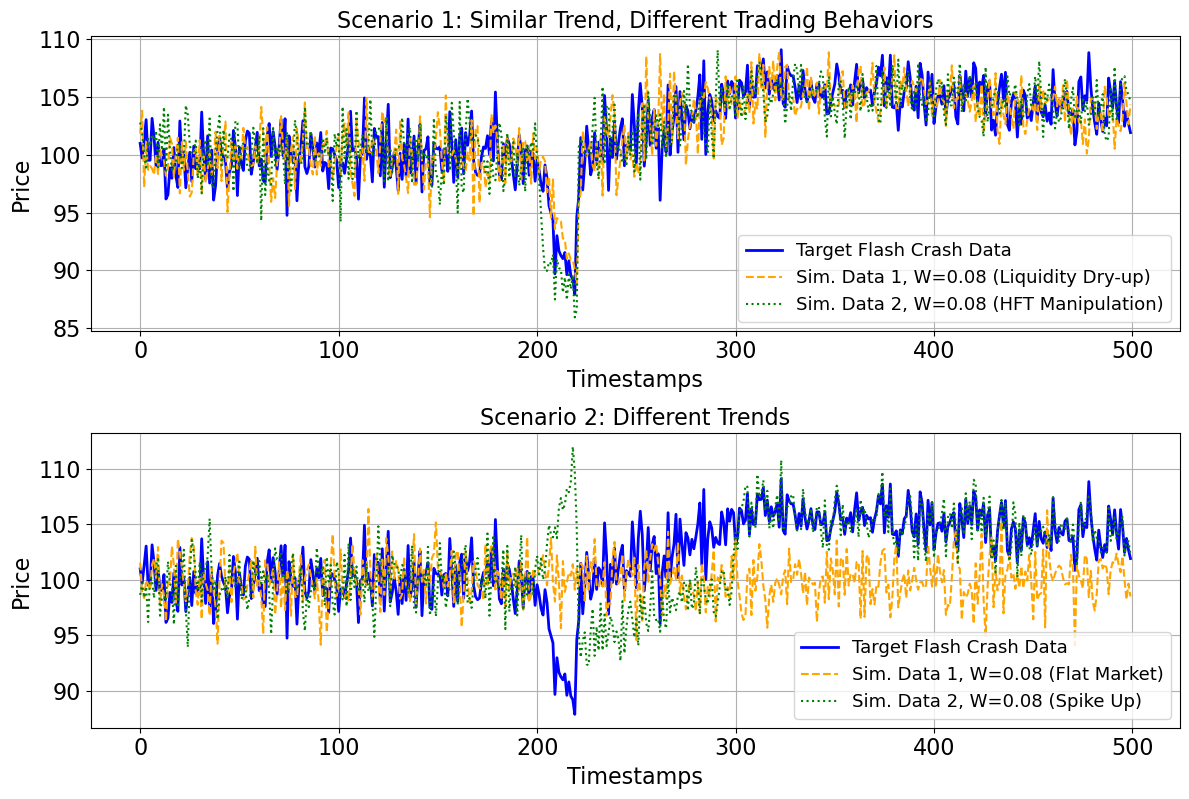}
  \caption{Two cases of non-identifiable simulation.}
  \label{figure:flash crash}
\end{figure}

Take the simulation of the well-known flash crush as an example. Two cases of non-identifiability are illustrated in Fig. \ref{figure:flash crash}. In case 1, the two calibrated parameters represent different trading behaviors while both two corresponding simulated data enjoy identical objective values. By analyzing the flash crush with the two parameters, one may get different causes (i.e., high liquidity taking with a large volume of market orders v.s. potential manipulation with a large volume of canceling orders). Also, the non-identifiability can lead to case 2, where two simulated data have the same discrepancy to the target data but appear to be totally different trends. In both cases, the calibration algorithm cannot identify the one with higher fidelity, which can deteriorate the downstream analysis.

Even though non-identifiability is frequently mentioned in the literature, there is neither a formal definition to describe it nor a rigorous way to alleviate it. This paper first defines non-identifiability as the probability of a randomly uniformly sampled parameter falling into the non-identifiable set. With this definition, this paper mathematically proves that the non-identifiability can be alleviated by calibrating with more distinct features, each of which represents an observed univariate time series. More specifically, we mathematically show that the upper bound of the non-identifiability can be reduced exponentially with the number of features if the selected features satisfy certain conditions. 

This explains well that the previous works mostly calibrate their models to merely one data feature and are thus less sufficient to describe the uniqueness of a social system. For clarification and applicability, this work restricts the scenarios to the FMS, as it is a representative social simulation and has been active for over 30 years \cite{palmer1994artificial}. Like other complex social systems, the real-world financial markets will output multivariate state data over time, e.g., price, volume, bid/ask directions, order arrival time, and other handcrafted features \cite{ntakaris2018benchmark}. However, existing works merely consider the observed mid-price data for calibration, resulting in wide criticism for their low simulation fidelity \cite{fagiolo2007critical}, especially in the high-frequency intraday simulations \cite{platt2018can,goosen2021calibrating}. 

Note that, following our theory, the task of calibrating to a multivariate time series is naturally divided into multiple individual tasks, each of which calibrates the model to one distinct univariate time series. In practice, how to jointly execute those individual tasks so that the obtained parameter is more identifiable to the original calibration tasks with multivariate time series? This challenges the implementation of the above theory. To address it, we propose a novel calibration objective by aggregating all those individual calibration tasks using the maximization function. This new objective function mathematically asks to search in the intersection of all those individual spaces. Thus, the obtained parameter can be highly identifiable across all individual tasks and thus fit our theory accurately. 

Extensive empirical studies have successfully shown that the proposed method can significantly alleviate the non-identifiability issue of FMS. Such advantages are verified robust over 6 commonly seen features of financial market data. Besides, this paper also presents the high-fidelity simulation results by using the proposed new calibration objective function on 10 synthetic data and 1 real data from the Shenzhen Stock Exchange of China. 

The remainder of this paper is as follows. Section~\ref{section:2} describes the background of FMS and related works. Section~\ref{section:3} theoretically discusses the alleviation of non-identifiability. Section~\ref{section:Empirical Studies} reports the empirical studies in detail. Section~\ref{section:conclusion} concludes this work.

\section{Background}\label{section:2}

\subsection{Calibration of Social Simulators}\label{section:2.2}
To ensure the simulation fidelity, the simulator often needs careful calibration of the agent parameters so that the simulated data resembles the observed time series from the real system. The general form of calibration refers to (\ref{eq:objective function}).

In recent years, the study of the discrepancy metrics $D$ has attracted increasing research attention, as the distance measure between time series remains unsolved and impacts significantly on the calibration performance. Methods of Simulated Moments \cite{grazzini2015estimation} is a typical method that first transforms the time series into several statistical moments and then calculates the weighted average between the moments of observed data and simulated data. The information criterion-based objective functions \cite{lamperti2018information,barde2017practical} abstract the temporal information of the original time series via various techniques like histograms and measure the distances between the abstracted temporal vectors. The non-parametric Kolmogorov-Smirnov test is also introduced as $D$ by statistically testing between the probabilistic distributions of the observed data and the simulated data \cite{bai2022efficient}. Other methods employ the Bayesian theory to estimate the likelihood \cite{kukacka2017estimation} or posterior distribution \cite{grazzini2017bayesian} of the observed data also require a well-defined $D$ for selecting effective samples and updating the estimated distributions.

\begin{figure}[tb]
  \centering
  \includegraphics[width=0.7\linewidth]{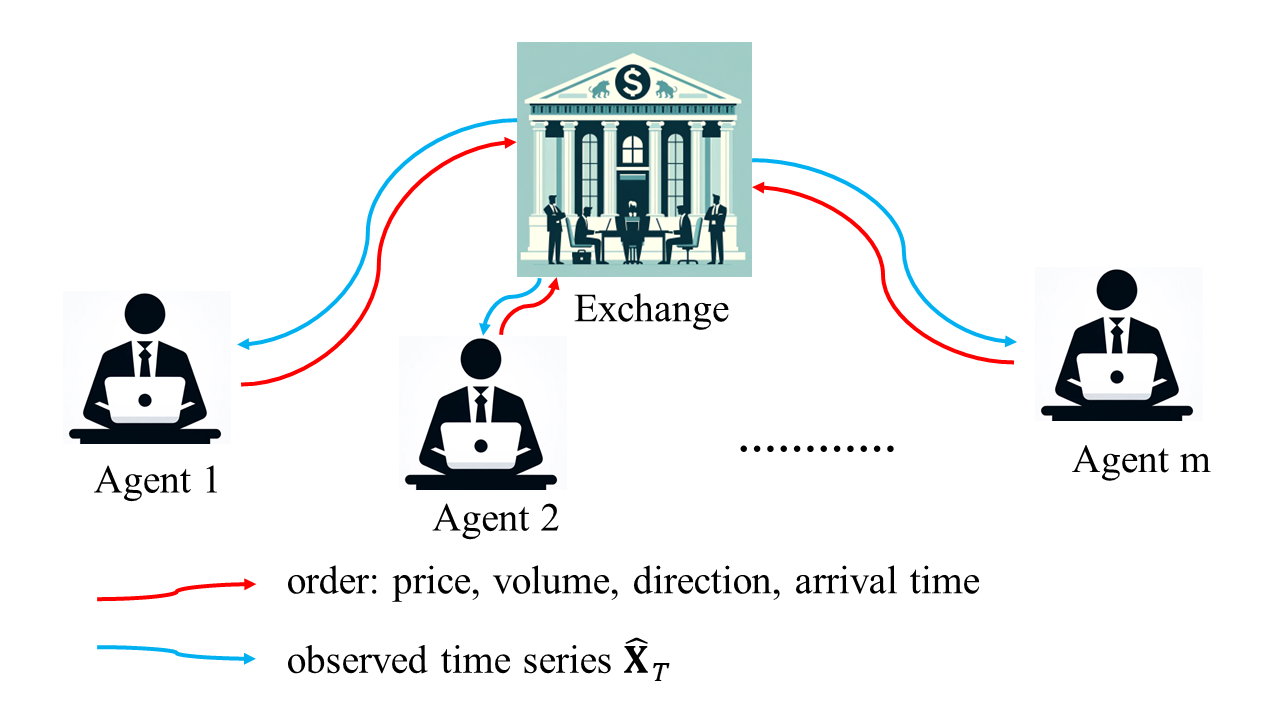}
  \caption{An illustration of the ABM-based FMS.}
  \label{figure:exchange and agent}
\end{figure}

Despite the progress of exploiting various metrics for calibrating social simulators, existing works frequently encounter the so-called non-identifiability issue, resulting in poor simulation fidelity \cite{lamperti2018agent}. From our viewpoint, one major reason is that existing works merely use one single feature to calibrate the model, ignoring that the real systems continuously output multivariate data. Although the features used might be a major factor in describing the social system, other features can help capture more details of the system. Unfortunately, existing calibration objectives are mostly designed for univariate time series data. Consequently, existing works are ill-equipped to calibrate with multivariate time series data. Yet, it is theoretically unclear how exactly it benefits from calibrating more features. 

\subsection{ABM-based FMS and Data Features}
In social simulations, ABMs are dominantly preferred over the black-box deep learning models due to their interpretability of the data-generating process \cite{jain2024limitorderbooksimulations,li2024simlob}. ABM constructs a collection of agents mimicking various social activities and interacting together to simulate the whole social system \cite{paulin2018agent}. In FMS, the agents are designed to model various trading strategies \cite{chen2012agent} (illustrated in Fig.~\ref{figure:exchange and agent}), e.g., fundamentalists, momentum traders, high-frequency traders, market makers, and arbitrage traders.

\begin{figure}[tb]
  \centering
  \includegraphics[width=0.7\linewidth]{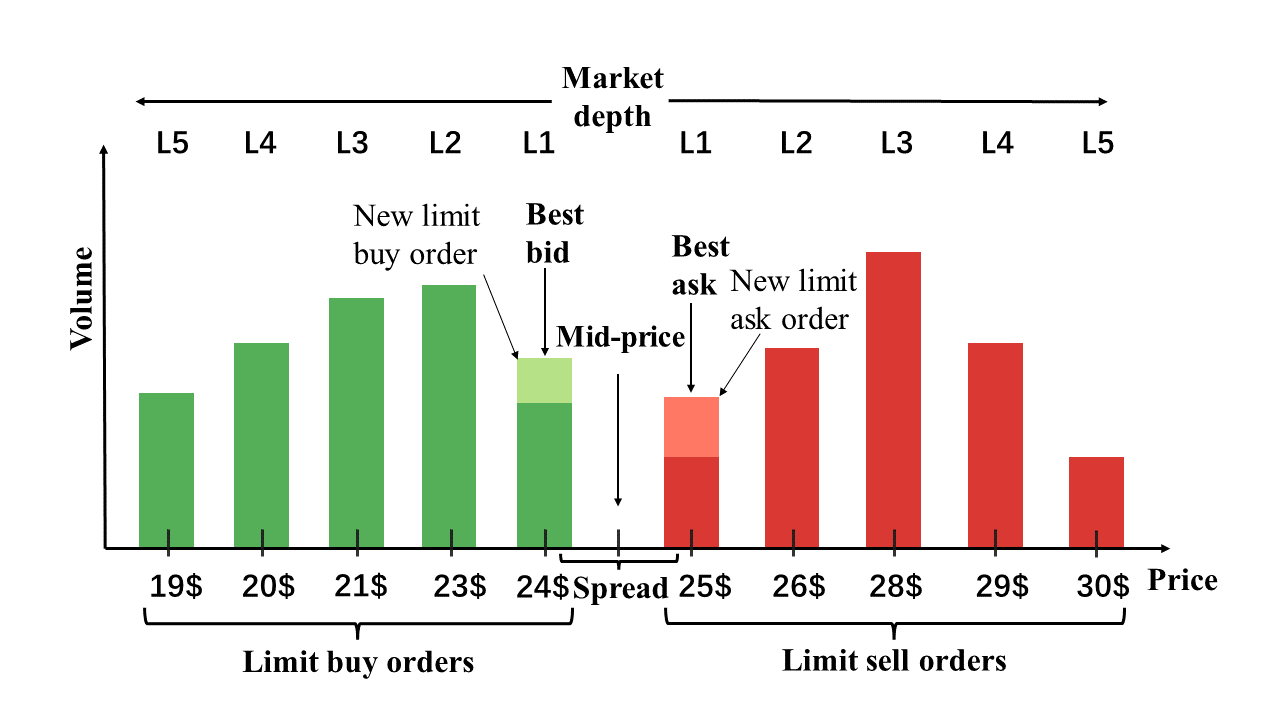}
  \caption{LOB contains much information about the market.}
  \label{figure:LOB}
\end{figure}

Regardless of the differences among agents, the ABM-based FMSs follow the same workflow \cite{axtell2025agent}. That is, at the beginning of the simulation, the parameterized agents are initialized with the calibrated parameters, which determine how exactly those agents submit orders under their own simulated strategies. Each submitted order usually consists of the price, volume, direction, and arrival time. The Limited Order Book (LOB), a particular data structure for organizing the existing untraded orders, is maintained by the exchange and can be observed by all trading agents. The simulated exchange generates the next LOB by matchmaking the incoming orders submitted by the agents and the current LOB \cite{vyetrenko2020get}.

Those untraded orders are categorized into two directions (bid or ask) and sorted by their prices. Orders with the same price are again sorted by arrival time. Each agent continuously observes the current LOB and submits its order at will. If a new order matches a price in the opposite direction of the LOB, it will be traded immediately, and the matched orders will be removed from the LOB. Undoubtedly, many data features can be extracted from LOB \cite{ntakaris2018benchmark}. However, existing FMSs mostly considered the mid-price as the only feature for calibration, which represents the average between the best ask price and best bid prices (see Fig.~\ref{figure:LOB}) and implies the potential price movement of the market.

Recent advances in financial time series modeling prioritize two objectives: capturing nonlinear dynamics while maintaining interpretability. The RHINE model \cite{xu2024rhine} addresses regime-switching through kernel-based spectral clustering and sliding-window eigengap analysis. Parallel innovations in concept drift detection, exemplified by WormKAN \cite{xu2024kan4drift}, employ Kolmogorov-Arnold Networks (KANs) with self-representation matrices to track abrupt transitions. These are unified through MT-KAN's \cite{xu2024kolmogorov} lightweight architectures for multivariate interaction modeling. Besides that, recent works like \cite{bai2022efficient} and \cite{li2024simlob} tried to calibrate simulators with multi-variate time series but did not investigate the non-identifiability issue from the perspective of using more features.

\section{A High-fidelity Calibration Objective}\label{section:3}
This section first formally defines the non-identifiability, then theoretically analyzes how it benefits from calibrating with multivariate data, and lastly proposes a new calibration objective using multiple features to implement our theory.

\subsection{The Definition of Non-identifiability}
As existing discrepancy metrics $D$ are designed for univariate data, we re-denote the target data as $\hat{\mathbf{X}}_T^k$ with a superscript $k$ to emphasize that it is a 1-dimensional time series describing some $k$-th feature of the real system. In financial markets, this feature can be either price, volume, bid/ask directions, order arrival time, or any other handcrafted features \cite{ntakaris2018benchmark}. Correspondingly, the $k$-th feature of the simulated data is re-denoted as $\mathbf{X}^k_T$. Let $\Omega^{D,k}$ be the parameter space of this calibration task. Then the original calibration task can be re-written as 

\begin{equation}\label{equation:calibration task}
\min _{\omega \in \Omega^{D,k}} D\left(\hat{\mathbf{X}}_T^k, M(\omega)=\mathbf{X}^k_T\right).   
\end{equation}

Without loss of generality, it is assumed that the optimal parameter $\omega^{*}$ lies in the defined parameter space, i.e., $\omega^{*}\in \Omega^{D,k}$, where $D(\hat{\mathbf{X}}_T^k, M(\omega^*))<\epsilon$. Note that the optimal parameter may not be unique by definition. The cardinality of the set containing all optimal parameters depends on 
$\epsilon$, which filters the observation noise in the data. This set, referred to as the non-identifiable set, is formally defined as follows:
\begin{definition}[\textbf{The Non-identifiable Set}]\label{def:set}
There exists some region $\mathcal{S}^{D,k,\epsilon} \subseteq \Omega^{D,k}$ that for all $ \omega_1, \omega_2 \in \mathcal{S}^{D,k,\epsilon}$ $D(\hat{\mathbf{X}}_T^k, M(\omega_1))<\epsilon$ and $D(\hat{\mathbf{X}}_T^k, M(\omega_2))<\epsilon$. Such region $\mathcal{S}^{D,k,\epsilon}$ is called the non-identifiable set.
\end{definition}

The Lebesgue measure (intuitively the hypervolume) of $\mathcal{S}^{D,k,\epsilon}$, denoted as $\mu\left(\mathcal{S}^{D,k,\epsilon}\right)$, is influenced by $\epsilon$. By choosing the parameter space to be finite and based on Definition~\ref{def:set}, we know that $0 < \mu\left(\mathcal{S}^{D,k,\epsilon}\right) \leq \mu\left(\Omega^{D,k,\epsilon}\right) < +\infty$. Then the non-identifiability can be defined as follows:

\begin{definition}[\textbf{The Univariate Non-identifiability}]\label{def:NIP_one_dimension}
Let $P\left(\omega \in \mathcal{S}^{D,k,\epsilon} \middle| \hat{\mathbf{X}}_T^k\right)$ be the non-identifiability to the $k$-th univariate time series data. It is defined as the probability of uniformly randomly sampling a parameter $\omega$ from $\Omega^{D,k}$ while it falls into $\mathcal{S}^{D,k,\epsilon}$. And we have 
\begin{equation}
P\left(\omega \in \mathcal{S}^{D,k,\epsilon}\middle|\hat{\mathbf{X}}_T^k\right) = \frac{\mu \left(\mathcal{S}^{D,k,\epsilon}\right)}{\mu \left(\Omega^{D,k}\right)}.
\end{equation}
\end{definition}

Therefore, the rank of any $\omega_1, \omega_2 \in \mathcal{S}^{D,k,\epsilon}$ cannot be effectively identified by the objective $D$ given the data $\hat{\mathbf{X}}_T^k$, leading to that the comparison-based calibration process easily gets stuck in $\mathcal{S}^{D,k,\epsilon}$. Note that, $\mathcal{S}^{D,k,\epsilon}$ can involve arbitrary parameters in $\Omega^{D,k}$, including the “ground-truth” parameter $\omega^{*}$.  Hence, if the calibration process converges to $\mathcal{S}^{D,k,\epsilon}$, one has to randomly pick one $\omega \in \mathcal{S}^{D,k,\epsilon}$ for further what-if simulation. Unfortunately, the underlying data-generating distribution decided by any randomly picked $\omega \in \mathcal{S}^{D,k,\epsilon}$ is less likely to be the same with $\omega^{*}$, given the nonlinear nature of the simulation model $M(\omega)$. Consequently, the what-if test on the randomly picked $\omega$ is less trustworthy unless the cardinality of $\mathcal{S}^{D,k,\epsilon}$ is sufficiently small.

\subsection{Alleviating the Non-identifiability with Multiple Features}
In this section, it is shown that the non-identifiability issue can be alleviated exponentially with $K>1$ features of the observed time series data.

Suppose the observed time series generated by the real social system is multivariate containing $K$ features, i.e., $\hat{\mathbf{X}}_T = \{\hat{\mathbf{X}}_T^k\}_{k=1}^K$, which is true for the financial markets. And the simulation model $M(\omega)$ can also generate a multivariate simulated data $\mathbf{X}_T=\{\mathbf{X}_T^k\}_{k=1}^K$ with the same $K$ features. Given that existing discrepancy metrics $D$ only compute for two univariate time series, there are naturally in total $K$ individual calibration tasks: $\min _w D(\hat{\mathbf{X}}_T^k, M(\omega)=\mathbf{X}_T^k), k=1, \cdots, K.$

Note that, given any two parameters $\omega_1, \omega_2 \in \Omega^{D,k}$, only they are simultaneously non-identifiable in all $K$ individual calibration tasks, i.e., $\omega_1, \omega_2 \in \{\mathcal{S}^{D,k,\epsilon}\}_{k=1}^K$, are they non-identifiable in the general calibration task with respect to multivariate $\hat{\mathbf{X}}_T$. Otherwise, their discrepancy to $\hat{\mathbf{X}}_T$ can be distinguished in at least one feature and thus are identifiable. The search space $\Omega^{D,k}$ is identical across all calibration tasks and is uniformly represented as $\Omega^D$. That is to say, only the intersection of the $K$ individual non-identifiable set $\{\mathcal{S}^{D, k,\epsilon}\}_{k=1}^K$ can be regarded as the non-identifiable set in the multivariate time series setting, which is defined as follows:

\begin{definition}[\textbf{The Multivariate Non-identifiability}]\label{def:NIP_multile_dimension}
Let $P\left(\omega \in\{\mathcal{S}^{D, k,\epsilon}\}_{k=1}^K \middle| \hat{\mathbf{X}}_T\right)$ be the non-identifiability to the multivariate time series with $K$ features. It is defined as the probability of uniformly randomly sampling a parameter $\omega$ from $\Omega^{D}$ while it falls into all $K$ individual non-identifiable set $\{\mathcal{S}^{D, k,\epsilon}\}_{k=1}^K$. And we have
\begin{equation}\label{equation:NIP_multile_dimension}
P\left(\omega \in\{\mathcal{S}^{D, k,\epsilon}\}_{k=1}^K \middle| \hat{\mathbf{X}}_T\right) = \frac{\mu \left( \cap_{k=1}^K \mathcal{S}^{D, k,\epsilon} \right)}{\mu \left( \Omega^D \right)}.
\end{equation}
\end{definition}

Since
\begin{equation}
\frac{\mu\left(\cap_{k=1}^K \mathcal{S}^{D, k, \epsilon}\right)}{\mu\left(\Omega^D\right)} \leq \min_{k} \frac{\mu\left(\mathcal{S}^{D,k,\epsilon}\right)}{\mu\left(\Omega^{D,k}\right)} \end{equation}
we have
\begin{equation}
P\left(\omega \in\{\mathcal{S}^{D,k,\epsilon}\}_{k=1}^K \middle| \hat{\mathbf{X}}_T\right) \leq \min_{k} P\left(\omega \in \mathcal{S}^{D,k,\epsilon}\middle|\hat{\mathbf{X}}_T^k\right).
\end{equation}

This equality holds only if the $K$ non-identifiable sets are fully overlapped. This can be caused by the fact that the $K$ features are fully dependent and that time series data of one feature can be deduced from other features. This suggests that the selected features should be as diverse as possible. Therefore, with proper choice of the features and $D$, the non-identifiability of calibrating the multivariate time series $\hat{\mathbf{X}}_T$ is smaller than that of the univariate $\hat{\mathbf{X}}_T^k$. In other words, calibrating multiple features can alleviate the non-identifiability issue. Next, we estimate how fast it decreases with $K$ as follows.

For clarity, let $A_k$ denote the event of a uniformly randomly sampled parameter falling into $\mathcal{S}^{D,k,\epsilon}$. The probability of happening $A_k$ is $P(A_k)=P(\omega \in \mathcal{S}^{D,k,\epsilon}|\hat{\mathbf{X}}_T^k)$. Then we have: 
\begin{subequations}
\begin{align}
&P\left(\omega \in\{\mathcal{S}^{D,k,\epsilon}\}_{k=1}^K \middle| \hat{\mathbf{X}}_T\right) \notag\\
&= P\left(A_1A_2 \cdots A_K\right) \notag \\[6pt]
&= P(A_1) \cdot P(A_2|A_1) \cdots P(A_K|A_{K-1}\cdots A_1) \notag \\[6pt]
&= \frac{\mu\left( \mathcal{S}^{D,1,\epsilon} \right)}{\mu\left( \Omega^{D} \right)} \cdot \frac{\mu\left( \bigcap_{i=1}^2 \mathcal{S}^{D,i,\epsilon} \right)}{\mu\left( \mathcal{S}^{D,1,\epsilon} \right)} \cdots \frac{\mu\left( \bigcap_{i=1}^K \mathcal{S}^{D,i,\epsilon} \right)}{\mu\left( \bigcap_{i=1}^{K-1} \mathcal{S}^{D,i,\epsilon} \right)}\\[6pt]
&= \beta_1 \cdot \beta_2 \cdot \beta_3 \cdots \beta_K \label{eq:line1} \\[6pt] 
&\leq \left(\frac{\sum_{i=1}^K \beta_i^2}{K}\right)^{\frac{K}{2}}. \label{eq:line2}\\\notag
\end{align}
\end{subequations}

The proof of (\ref{eq:line2}) is provided in appendix \ref{appendix:proof1}.
In (\ref{eq:line1}), $\beta_1$ represents the non-identifiability of feature 1, i.e., the probability that any $\omega \in \mathcal{S}^{D,1,\epsilon}$ falls into $\Omega^{D}$. $\beta_i, i=2,...,K$ denotes the probability that any parameter $\omega \in \cap_{k=1}^{i}\mathcal{S}^{D,k,\epsilon}$ falls into $\cap_{k=1}^{i-1}\mathcal{S}^{D,k,\epsilon}$. In other words, $\beta_i$ means the overlapping ratio between $\mathcal{S}^{D,i,\epsilon}$ and $\cap_{k=1}^{i-1} \mathcal{S}^{D,k,\epsilon}$, and $\beta_i \leq 1$. Especially, $\beta_i = 1$ only if $\cap_{k=1}^{i-1} \mathcal{S}^{D,k,\epsilon} \subseteq \mathcal{S}^{D,i,\epsilon}$, indicating that the selected $i$-th feature is fully dependent on the first $i-1$ features.

The order of features in (\ref{eq:line1}) can be arbitrary, but it influences the upper bound of the multivariate non-identifiability, which is defined as (\ref{eq:line2}). That is,

\begin{theorem}[\textbf{The Exponential Alleviation}]\label{def:expall}
If each $K$-th feature is selected such that $\beta_{K}^{2} < \frac{{\sum}_{i=1}^{K-1}\beta_i^2}{K-1}$, the upper bound of the multivariate non-identifiability is reduced exponentially with the increasing number of features $K$.
\end{theorem}

This is intuitive that if $\beta_{K}^{2} < \frac{{\sum}_{i=1}^{K-1}\beta_i^2}{K-1}$, the base of (\ref{eq:line2}) will not increase.
This theorem not only provides an exponential decreasing upper bound for the multivariate non-identifiability but theoretically suggests a rule to select the features. The proof of Theorem~\ref{def:expall} is in appendix \ref{appendix:proof2}.

\subsection{Aggregating Multiple Features via Maximization}
Note that the above discussions require the multivariate time series $\{\hat{\mathbf{X}}_T^k\}_{k=1}^K$ being calibrated separately as $K$ univariate time series with existing discrepancy metrics $D$. This results in $K$ individual calibration tasks of $D(\hat{\mathbf{X}}_T^k,M(\omega)=\mathbf{X}_T^k), k = 1, \cdots, K$. How to jointly calibrate the parameter to satisfy the $K$ individual tasks simultaneously? This section proposes a new objective function based on the $K$ individual objectives so that the practical alleviation of the non-identifiability follows the above theory. 

\begin{figure}[h]
  \centering
  \includegraphics[width=1\linewidth]{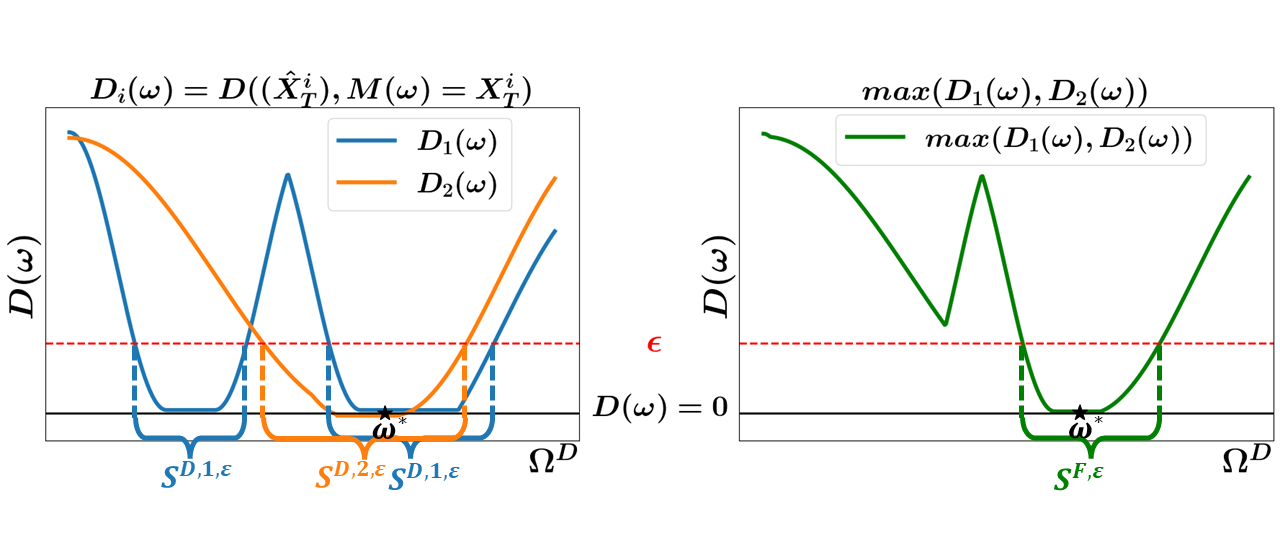}
  \caption{Optimizing the maximization of two functions equals searching in their intersections.}
  \label{figure:intersection=max}
\end{figure}

Let $F$ be the proposed new objective function used for calibrating the multivariate time series $\hat{\mathbf{X}}_T$, and let $\mathcal{S}^{F,\epsilon}$ denote the associated non-identifiable set defined by this objective function. 
To achieve $\mathcal{S}^{F,\epsilon} = \cap_{k=1}^K \mathcal{S}^{D,k,\epsilon}$, the proposed way is to ensure every location in $S^{F}$ to be the worst value among all $K$ non-identifiable sets $\{\mathcal{S}^{D,k,\epsilon}\}_{k=1}^K$ (see Fig.~\ref{figure:intersection=max} for illustration). In this case, any parameter $\omega$ that is not simultaneously in all $K$ non-identifiable sets will make it fall outside of the $\mathcal{S}^{F,\epsilon}$. This is because the parameters outside the non-identifiable set are even worse than those that are non-identifiable from the ground-truth parameter $\omega^*$.

Given that the calibration problem is a minimization problem, the new objective function $F$ can be implemented with a maximization aggregation among $K$ individual calibration tasks with $D$. Formally, we have the following theorem.
\begin{theorem}(\textbf{The utility and  uniqueness of new objective function $F$})\label{thm:max=intersection}
Let $F$ aggregate the $K$ individual calibration tasks via maximization, i.e.
\begin{equation}\label{eq:F_objective_function}
\begin{aligned}
&F\left(\hat{\mathbf{X}}_T,M(\omega)=\mathbf{X}_T\right) \\
=&\max _{k \in \{1,\cdots,K\}} D\left(\hat{\mathbf{X}}_T^k,M(\omega)=\mathbf{X}_T^k\right),
\end{aligned}
\end{equation}
then minimizing $F$  uniquely achieves $\mathcal{S}^{F,\epsilon} = \cap_{k=1}^K \mathcal{S}^{D,k,\epsilon}$.
\end{theorem}

The proof of Theorem~\ref{thm:max=intersection} is provided in appendix \ref{proof:uiqueness_of_max}.

% Furthermore, setting $\epsilon = 0$ corresponds to the theoretical scenario, affirming that the discussions in this section are applicable under both theoretical and practical conditions.

\section{Empirical Studies}\label{section:Empirical Studies}
This section empirically verifies two major claims. 
\begin{itemize}
\item With the increasing number of features applied in the calibration, the non-identifiable set shrinks exponentially.
\item The fidelity of the simulation data, in terms of various performance metrics, is improved significantly by calibrating more features.
\end{itemize}

\begin{table*}[t]
\caption{The ranges for randomly sampling parameters $\omega$ to generate the synthetic data}
\label{table:The ranges for randomly sampling parameters}
\centering
\begin{tabular}{c||cl} \hline\hline
\textbf{Para.} & \textbf{Range} & \textbf{Remarks} \\\hline\hline
$\delta$ &$[0.00,0.050]$ & The probability for each liquidity taker to cancel an untraded order.\\
$\lambda_0$ &$[50.00,300.000]$ & Controlling the price of each limited order.\\
$C_{\lambda}$ &$[1.00,50.000]$ & Controlling the price of each limited order.\\
% $\Delta_S$  &$[0.0005,0.005]$ & Controlling the probability of the side of a market order (ask or bid).\\
$\Delta_S$  &$[0.00,0.005]$ & Controlling the probability of the side of a market order (ask or bid).\\
$\alpha$ &$[0.05,0.500]$ & The probability of each liquidity provider submitting a limited order.\\
$\mu$ &$[0.00,0.050]$ & The probability of each liquidity taker submitting a market order.\\
\hline\hline 
\end{tabular}
\end{table*}

\subsection{Experimental Settings}
To conduct a calibration process, 4 terms should be specified according to (\ref{eq:F_objective_function}), i.e., the discrepancy metric $D$, the simulation model $M(\omega)$, the optimization algorithm $\rm{min}_\omega$, and the $K$ features $\{\hat{\mathbf{X}}_T^k\}_{k=1}^K$.

\textbf{The discrepancy metric}. The Wasserstein distance is employed as $D$ since it provides geometric interpretability and tail sensitivity \cite{coletta2021towards, coletta2022learning}. Given two univariate time series $\mathbf{X}_T^k$ and $\hat{\mathbf{X}}_T^k$, the Wasserstein distance is defined as
\begin{equation}
\begin{aligned}
&D(\hat{\mathbf{X}}_T^k, \mathbf{X}_T^k) \\
=&\int_{-\infty}^{+\infty}\left|\frac{1}{T} \sum_{t=t_s}^{t_e} \mathcal{I}(\hat{\mathbf{X}}_t^k \leq x)-\frac{1}{T} \sum_{t=t_s}^{t_e} \mathcal{I}(\mathbf{X}_t^k \leq x)\right| \rm{d}\textit{x},
\end{aligned}
\end{equation}

\noindent where $\mathcal{I}(\cdot)$ is an indicator function that returns 1 if the input event is true. Otherwise, it returns 0. 

\textbf{The simulation model.} This work considers the Preis-Golke-Paul-Schneid (PGPS) model as the simulation model due to its popularity in recent FMS works \cite{platt2018can,goosen2021calibrating}. The PGPS model simulates the market dynamics through interactions between two types of agents, i.e., 125 liquidity providers and 125 liquidity takers. The PGPS model has 6 hyper-parameters shared by all agents that need to be calibrated, i.e., $\omega = [\delta, \lambda_0, C_{\lambda}, \Delta_s, \alpha, \mu]$. At the $t$-th time-step, each liquidity provider submits a limited order at a fixed probability $\alpha$ with the default volume equals 1. The probability of a limited order being bid side or ask side equals 0.5. Each liquidity taker submits a market order at a fixed probability $\mu$ with the default volume equals 1. The probability of a market order being bid side or ask side is $q_{taker}(t)$ or $1-q_{taker}(t)$, respectively. The probability $q_{taker}(t)$ is specified by a mean-reverting random walk with mean equals 0.5, mean reversion probability equals $0.5 + |q_{taker}(t) - 0.5| $, and the increment sizes towards mean equals $\pm \Delta_s$. Moreover, the liquidity taker has a probability $\delta$ to cancel its untraded limited order. Let $p_a(t)$ and $p_b(t)$ represent the best ask and bid price, respectively. The price of a market order is determined by the market, i.e., the best price at the opposite side of the LOB. The price of a limit order is determined as $p = p_s(t) + \lambda(t) \log(u) + s$ where $p_s(t) = p_a(t)$, $s = 1$ for an ask order, and $p_s(t) = p_b(t)$, $s = -1$ otherwise, with $u \sim U(0,1)$. Here $\lambda(t)$ is a time-variant order placement depth parameter and is calculated as
\begin{equation}
\lambda(t)=\lambda_0\left(1+\frac{\left|q_{taker}(t)-\frac{1}{2}\right|}{\sqrt{<q_{taker}(t)-\frac{1}{2}>^2}} C_\lambda\right) 
\end{equation}

\noindent $\sqrt{<q_{taker}(t)-\frac{1}{2}>^2}$ is a pre-computed value. Before initiating the simulation, it is computed in advance through $10^{5}$ Monte Carlo iterations, ensuring its convergence to the accurate standard deviation.

\textbf{The optimization method.} Since the order matchmaking process needs to align with the real market, the PGPS model should be run on the executable trading programs. As a result, the calibration objective function is non-differentiable and the calibration process often has to resort to non-derivative optimization methods. In such contexts, both Bayesian optimization methods and evolutionary algorithms are particularly suitable. This work employs the well-known Particle Swarm Optimization (PSO) algorithm \cite{kennedy1995particle}, which has been frequently adopted to train the financial simulation or prediction models\cite{platt2020comparison,li2024simlob,bazrkar2023predict}, to optimize $\omega$ with respect to (\ref{eq:F_objective_function}). For simplicity, the standard version of PSO \cite{shi1998modified} is considered. The pseudocode for calibration with PSO is shown in Algorithm~\ref{alg:algorithm}. 

\begin{algorithm}[H]
\caption{Calibration with PSO}
\label{alg:algorithm}
\textbf{Input}: Number of Iterations to run $\kappa$\\
\textbf{Parameter}: The hyper-parameters of PSO\\
\textbf{Output}: The optimal candidate $\omega^{*}$

\begin{algorithmic}[1] %[1] enables line numbers
\STATE Let $t=0$.
\STATE Initialize a population of $N$ particles $\{\omega_i\}^N_{i=1}$.
\STATE Simulate each of $N$ particles with $M(\omega)$ to obtain $N$ simulated data.
\STATE Calculate the performance of each of $N$ particles as the Wasserstein distance.
\STATE Set $\omega^*$ to the best performed candidate.
\WHILE{$t<\kappa$}
\STATE Generate a new population of $N$ particles $\{\omega_i\}^N_{i=1}$ using PSO operators.
\STATE Simulate each of $N$ particles with $M(\omega)$ to obtain $N$ simulated data.
\STATE Calculate the performance of each of $N$ particles as the Wasserstein distance. 
\STATE Update $\omega^*$ to the best performed candidate.
\ENDWHILE
\STATE \textbf{return} $\omega^*$
\end{algorithmic}
\end{algorithm}

The PSO hyper-parameters, population size $N$, inertia weight $\omega$, cognitive/social coefficients $c_1$, $c_2$ are set to $N=40$, $\omega=0.8$, $c_1 = c_2 = 0.5$, chosen from ranges $N \in [30, 50]$, $\omega \in [0.6, 1.0]$, $c_1, c_2 \in [0.3, 0.7]$ based on experiments.

% Without further fine-tuning, the hyper-parameters of PSO follow the suggested configurations, where the population size is set to $N=40$, the inertia weight is set to 0.8, the cognitive and social crossover parameters $c_1 = 0.5, c_2 = 0.5$.

\begin{table}[t]
\caption{Parameter settings of generating 10 synthetic data with the PGPS model}
\label{table:Parameter settings of generating 10 synthetic data with the PGPS model}
\centering
    \begin{tabular}{r||rrrrrr}\hline\hline
    \textbf{Parameters}  & \textit{$\mathbf{\delta}$} & \textit{$\mathbf{\lambda_0}$} &  \textit{$\mathbf{C_{\lambda}}$} & \textit{$\mathbf{\Delta_S}$} & \textit{$\mathbf{\alpha}$} &  \textit{$\mathbf{\mu}$}   \\ \hline\hline
        data 1 & 0.025 & 100 & 10 & 0.001 & 0.15 & 0.025\\
        data 2 & 0.002 & 200 & 10 & 0.002 & 0.1 & 0.03\\
        data 3 & 0.05 & 152 & 45 & 0.003 & 0.13 & 0.05\\
        data 4 & 0.02 & 288 & 27 & 0.003 & 0.11 & 0.04\\
        data 5 & 0.04 & 62 & 35 & 0.003 & 0.12 & 0.04\\
        data 6 & 0.01 & 171 & 15 & 0.0015 & 0.15 & 0.03\\
        data 7 & 0.033 & 114 & 14 & 0.0033 & 0.1 & 0.047\\
        data 8 & 0.01 & 129 & 30 & 0.0017 & 0.05 & 0.03\\
        data 9 & 0.02 & 62 & 2 & 0.003 & 0.14 & 0.02\\
        data 10 & 0.01 & 87 & 23 & 0.001 & 0.09 & 0.05\\
        \hline\hline
\end{tabular}
\end{table}

\textbf{The selected features.} Six different yet commonly seen features are considered \cite{vyetrenko2020get,cont2001empirical}, which mainly concern either the price or volume information of the LOB data and are listed as follows:
\begin{itemize}
    \item $f_1$, the mid-price $m_{t}=\frac{p_a(t)+p_b(t)}{2}$ at each $t$-th step;
    \item $f_2$, the total traded volume within each $t$-th step;
    \item $f_3$, the price return at each $t$-th step $\ln m_{t+1} - \ln m_{t}$;
    \item $f_4$, the spread at each $t$-th step $p_a(t)-p_b(t)$;
    \item $f_5$, the volume of the best bid price at each $t$-th step;
    \item $f_6$, the volume of the best ask price at each $t$-th step.
\end{itemize}

By incorporating the 6 selected features, 6 objective functions are constructed based on (\ref{eq:F_objective_function}), denoted as $F_i, i=1,2,...,6$. Each $F_i$ indicates that the first $i$ features are used for calibration, i.e., $F_i=\max _{k \in  \{1,\cdots,i\}  } D(\hat{\mathbf{X}}_T^k,M(\omega))$. The following empirical studies show how the increasing number of features influences the simulation fidelity.

\textbf{The test protocol.} The test data consists of 11 targeted time series to be calibrated, including 10 synthetic data sets and 1 real data set. To ensure sufficient diversity, the 10 synthetic time series are generated using the PGPS model by randomly sampling parameters within the ranges specified in table~\ref{table:The ranges for randomly sampling parameters}, as suggested by \cite{platt2018can}. The 10 sampled parameters are listed in table~\ref{table:Parameter settings of generating 10 synthetic data with the PGPS model}. Each synthetic time series comprises 3600 time steps at a frequency of 1 second. The real data is chosen as 000001.sz from Shenzhen Stock Exchange of China, consisting of 1200 time steps at a frequency of 3 seconds from 9:30 a.m. to 10:30 a.m. of a day in 2019. The time budget of $10^4$ simulation evaluations is allowed for the calibration of each targeted data. That is, the PSO runs for 250 iterations with a population size of 40, and the best parameter with the lowest objective function value is applied to the PGPS model to simulate data as $\mathbf{X}_T$.

\begin{figure}[t]
    \centering
     \includegraphics[width=\linewidth]{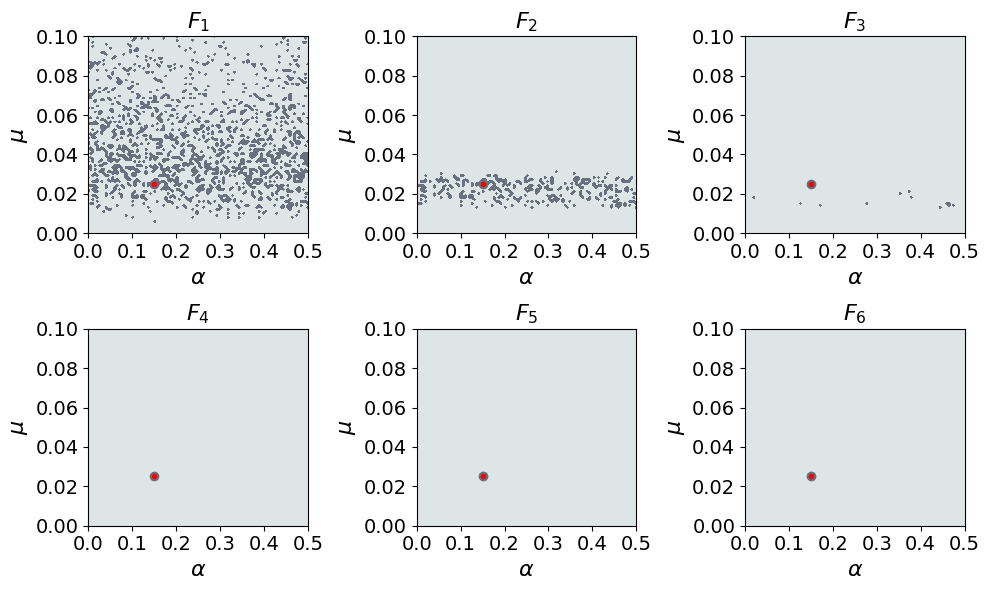} 
    \caption{The illustration of the exponentially decreased non-identifiable set (grey dots). The red dot is the optima.}
    \label{figure:multiple features}
\end{figure}

\begin{figure}[t]
    \centering
     \includegraphics[width=0.7\linewidth]{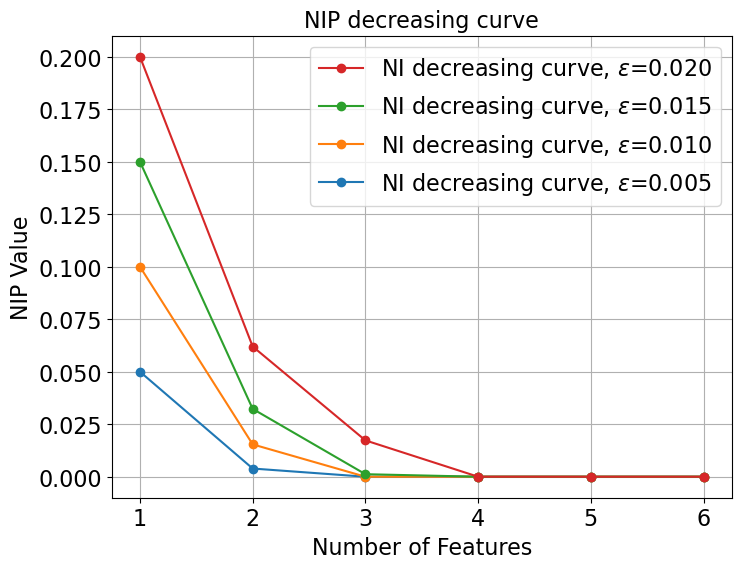} 
    \caption{The non-identifiability (NI) decreases exponentially with the increasing features.}
    \label{figure:multiple_curve}
\end{figure}

\begin{figure}[t]
  \centering
  \includegraphics[width=0.7\linewidth]{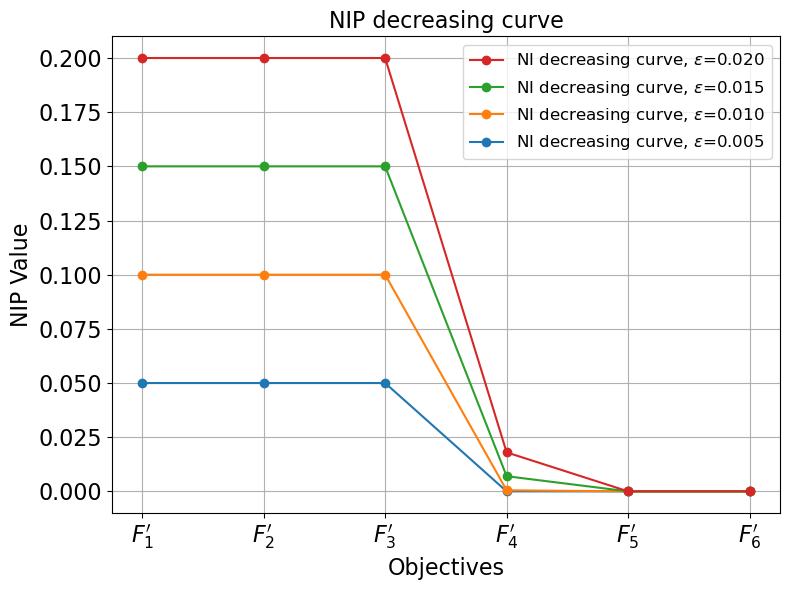}
  \caption{The non-identifiability (NI) decreasing behaviors with the addition of new features.}
  \label{figure:New_NIP}
\end{figure}

Calibration performance is evaluated using the Wasserstein (W) distance and Mean Square Error (MSE), two common metrics in time series analysis. Both metrics measure the average discrepancy between synthetic and simulated data across 6 features, i.e., $\frac{\sum_{k=1}^{6}D(\hat{\mathbf{X}}_T^k, M(\omega)=\mathbf{X}_T^k)}{6}$, where $D$ denotes W or MSE. Each feature is normalized by the min-max method.

\textbf{The computational platform.} This work builds the simulation environment based on the Multi-Agent Exchange Environment (MAXE) simulator developed by the University of Oxford \cite{belcak2021fast}. The experiments were run on a server with 500GB of memory, the Intel(R) Xeon(R) Platinum 8358 CPU @ 2.60GHz with 32 physical cores. For a single objective, PSO can be run in parallel across 40 simulators. The calibration of 6 objective functions for 10 synthetic data takes 48 to 60 hours in a data-parallel manner using the CPU.

\subsection{Results on Alleviation of Non-identifiability} Fig.~\ref{figure:multiple features} illustrates the changes of the non-identifiable sets in the 2-dimensional parameter space of $[\alpha, \mu]$. This is done by three steps: 1) a target data of 3600 time steps is generated using the PGPS model with the recommended default parameter setting $\omega^*=[0.025,100,10,0.001,0.15,0.025]$ \cite{goosen2021calibrating}; 2) On this default setting, 10000 combinations are created by sampling in $[\alpha, \mu]$ in the grid while keeping the values of other 4 parameters unchanged; 3) By using $F$ to measure the discrepancy between the target data and the simulated data of each of the 10000 combinations, it obtains the objective values of the 10000 combinations on each $F_i$, which are depicted as Fig.~\ref{figure:multiple features}. The 6 figures in Fig.~\ref{figure:multiple features} are depicted according to the objective values of $F_1$, $F_2$,...,$F_6$, respectively. Only the combinations whose objective values are smaller than $\epsilon = 0.1$ are depicted in grey; otherwise, they will be depicted in blue. For example, in the third figure marked with $F_3$, the combinations depicted in grey are those whose discrepancy values are smaller than $\epsilon = 0.1$ on all the first 3 features $\{f_i\}_{i=1}^3$. By this means, the grey points indicate the non-identifiable set as they are indistinguishable from $\omega^*$ (marked as the red dot). 
As can be seen, with the increase of the features, the grey non-identifiable set decreases significantly. When 4 are incorporated in $F$, only the optimum exists, indicating that the non-identifiability issue is indeed alleviated very effectively. 

Fig.~\ref{figure:multiple_curve} further depicts how the non-identifiability decreases with the increasing numbers of features with different settings of $\epsilon$. This is done by calculating (\ref{eq:line1}) using the above 10000 combinations.
Clearly, all the curves follow an exponentially decreasing manner. Besides, a larger value of $\epsilon$ is, the slower the corresponding curve decreases. 

\begin{figure}[t]
  \centering
  \includegraphics[width=1\linewidth]{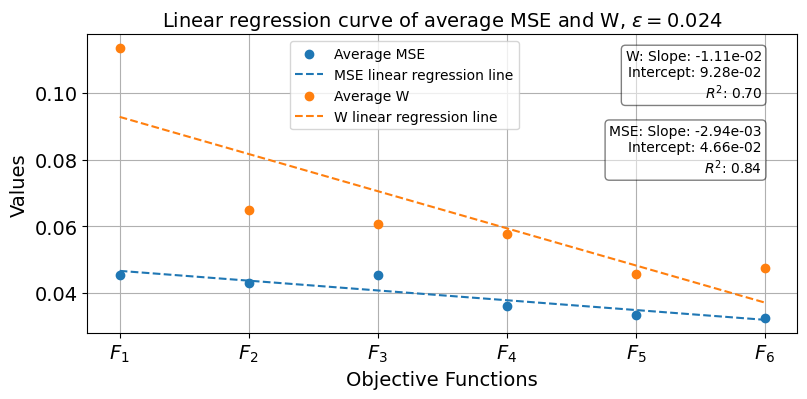}
  \caption{The distances of the simulated data to the target data decrease almost linearly with the increasing number of features incorporated in $F$.}
  \label{figure:multipleepsilon}
\end{figure}

\begin{figure}[b]
  \centering
  \includegraphics[width=0.6\linewidth]{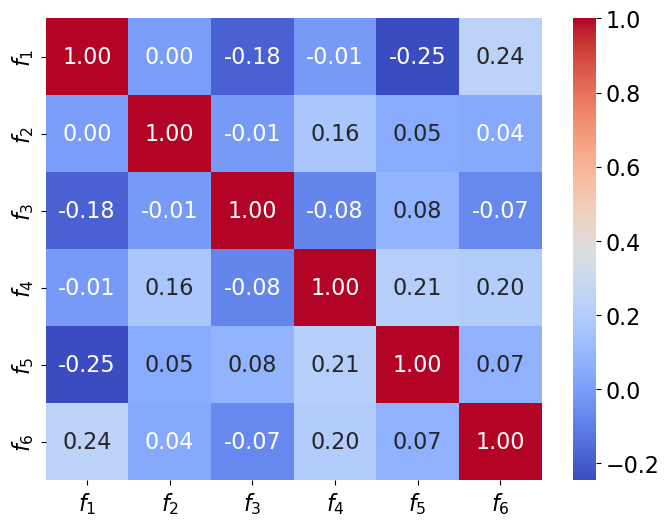}
  \caption{The average Pearson correlation coefficient of the selected features among 10 simulated data.}
  \label{figure:heatmap}
\end{figure}

Building on this, an additional experiment investigates the impact of feature dependencies on the alleviation. The theoretical proof in Appendix \ref{proof:uiqueness_of_max} suggests that linearly dependent features do not contribute to the alleviation. To validate this, six new test features are constructed based on the original six, where $f_1^{\prime} = f_1, f_2^{\prime} = 2f_1, f_3^{\prime}=-2f_1, f_4^{\prime} = f_5, f_5^{\prime} = f_6, f_6^{\prime} = f_2$.  The corresponding new objectives, $F_i^{\prime}$, are constructed in the same manner as the original $F_i$.
These features include both linearly dependent ones ($f_1^{\prime}, f_2^{\prime}, f_3^{\prime}$) and non-linearly related ones ($f_4^{\prime}, f_5^{\prime}, f_6^{\prime}$), enabling a direct evaluation of their respective effects on NIP reduction.
As illustrated in Fig. \ref{figure:New_NIP}, adding linearly dependent features does not reduce non-identifiability, aligning with the theoretical findings in Appendix \ref{proof:uiqueness_of_max}. In contrast, introducing non-linearly related features leads to a significant reduction, confirming that independent features impose additional constraints on the non-identifiability set.

While adding more independent features continues for alleviation, the marginal benefit diminishes as more features are included. This effect arises because the initial features already contribute substantially to the reduction, leaving less room for further alleviation. As a result, even if newly added features are entirely independent or possess highly distinctive and informative properties, their impact on further shrinking the non-identifiability set is inherently limited. Mathematically, this corresponds to their associated $\beta$ values approaching 1, reflecting the decreasing marginal effect of additional features rather than an indication of feature redundancy.

To further assess the effectiveness of the originally selected six features, the average Pearson correlation coefficient among ten target datasets is computed, as shown in Fig.~\ref{figure:heatmap}. 
The results indicate low pairwise correlations among the selected features, supporting their effectiveness in reducing NIP and confirming their direct applicability in calibration.

\begin{figure*}[t]
  \centering
  \includegraphics[width=1\linewidth]{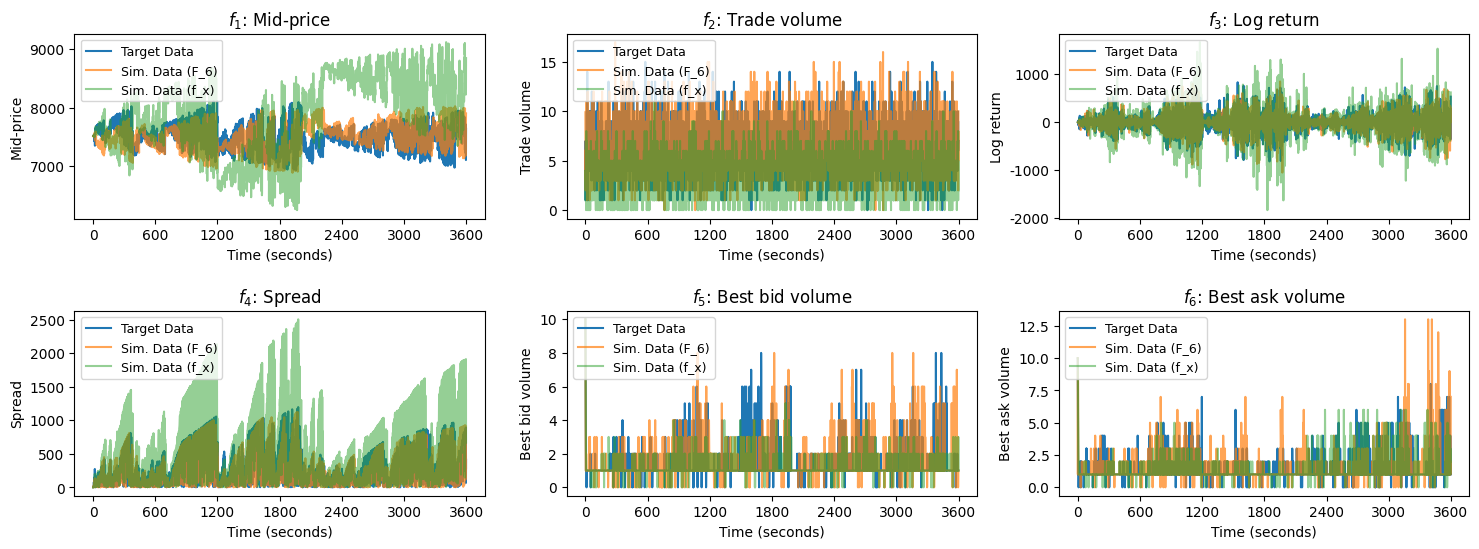}
  \caption{The simulated data using $F_6$ resembles the synthetic data on all 6 features.}
  \label{figure:simulationdata}
\end{figure*}

\begin{table}[t]
\centering
\begin{threeparttable}
\caption{Comparisons of calibration performance between 6 univariate data and $F_6$.\tnote{*}}
\label{table:synthetic data}
\begin{tabular}{@{}c|cccccc|c@{}}
\toprule
\multicolumn{1}{c|}{} & $f_1$ & $f_2$ & $f_3$ & $f_4$ & $f_5$ & $f_6$ & $F_6$\\ 
\midrule
\multirow{1}{*}{Data 1}  & 3.99 & 5.07 & 6.22 & 4.62 & 4.92 & 3.95 & \textbf{3.07}\\
\midrule
\multirow{1}{*}{Data 2}  & 3.65 & 3.95 & 4.67 & 3.73 & 6.18 & 4.77 & \textbf{2.90}\\
\midrule
\multirow{1}{*}{Data 3}  & \textbf{2.86} & 4.43 & 3.73 & 3.87 & 5.75 & 4.62 & 3.08\\
\midrule
\multirow{1}{*}{Data 4}  & 2.83 & 3.50 & 3.47 & 3.87 & 4.33 & 2.99 & \textbf{2.47}\\
\midrule
\multirow{1}{*}{Data 5}  & 2.72 & \textbf{2.70} & 4.93 & 5.98 & 3.17 & 5.83 & 2.84\\
\midrule
\multirow{1}{*}{Data 6}  & 5.23 & 5.58 & 4.02 & 6.67 & 4.48 & 6.77 & \textbf{3.67}\\
\midrule
\multirow{1}{*}{Data 7}  & 3.08 & 3.57 & \textbf{3.06} & 3.65 & 3.60 & 3.86 & 3.51\\
\midrule
\multirow{1}{*}{Data 8}  & 4.88 & 4.50 & 2.56 & 3.82 & 3.91 & 3.62 & \textbf{2.26}\\
\midrule
\multirow{1}{*}{Data 9}  & 4.31 & 3.59 & 5.07 & 4.19 & 5.73 & 3.82 & \textbf{3.51}\\
\midrule
\multirow{1}{*}{Data 10} & 2.53 & 2.43 & 2.95 & 6.93 & 3.22 & 2.24 & \textbf{2.23}\\
\midrule
\multirow{1}{*}{\#rank}  & $3.20^{+}$ & $4.10^{+}$  & $4.00^{+}$  & $5.10^{+}$  & $5.60^{+}$  & $4.50^{+}$  & \textbf{1.50} \\
\bottomrule
\end{tabular}
\begin{tablenotes}
\footnotesize
% \item[*] Note that the performance is at the $1E10^{-2}$ level. Here + denotes statistical significance at the $\alpha = 0.05$ level ($p<0.05$).
\item[*] Performance is at the $10^{-2}$ level. $+$ denotes significance under Bonferroni correction (family-wise $\alpha = 0.05$; $m = 6$, per-test $\alpha = 0.0083$) in pairwise comparisons between $F_6$ and $f_1,\cdots,f_6$.
\end{tablenotes}
\end{threeparttable}
\end{table}

\subsection{Results on Calibration to the Synthetic Data} 
Note that although the non-identifiable set of any $F_i$ is verified to be smaller than that of any $F_j$, $i>j$, it does not deterministically lead to better calibration performance of $F_i$ than $F_j$. Because $F$ mainly decides the parameter space, while the calibration performance is also jointly influenced by the optimization algorithms. Given the calibration problem is black-box with multiple local optima, it is non-trivial for the optimization algorithms to find the global optimum even if the parameter space is modeled fully identifiable. 

For each of the 10 synthetic data, by optimizing the 6 objective functions $F_i, i=1,...,6$ with PSO, it results in 6 calibrated models and 6 simulated data. Note that, even though a model is calibrated using $F_i, i<6$, it can still functionally generate the simulated data $\mathbf{X}_T$ with respect to all 6 features.  

The performance between each simulated data and its corresponding synthetic data is measured by the indicators based on both MSE and W, respectively. In Fig.~\ref{figure:multipleepsilon}, each (blue or yellow) point displays the average performance on these 10 synthetic data. It shows that by using more features in $F$, the calibration performance generally improves as the distance values decrease.  

Furthermore, it is interesting to see how the simulation model calibrated by all features ($F_6$) compares to the ones calibrated to every single feature $f_i, i=1,...,6$, instead of in an incremental manner of $F_i, i=1,...,5$. As listed in Table~\ref{table:synthetic data}, by calibrating with $F_6$, the obtained parameter enjoys the best simulation performance, i.e., the smallest averaged distances on 7 out of the 10 instances and the top 3 averaged distances on the other 3 instances. Besides, the Friedman test ($p = 6.7 \times 10^{-4}$) indicates significant ranking differences among methods. Subsequent pairwise Wilcoxon signed-rank tests, with Bonferroni correction, confirm that $F_6$ consistently outperforms $f_1$ to $f_6$ with statistical significance. This implies that the proposed theory and new objective function are applicable to various features and not restricted by the feature with more information about the social system, e.g., the mid-price.

We further expand the experiments in Fig.~\ref{figure:multipleepsilon} to more different values of $\epsilon$. Note that $\epsilon$ determines how the random noise in the financial market data impacts the measure of the discrepancy between two time series. The higher the value of $\epsilon$ is, the more likely two time series are indistinguishable given Definition ~\ref{def:set}. 
The results are listed in Table~\ref{table:differentepl}. It can be seen that under different $\epsilon$, the calibration performance always decreases linearly with the increasing number of features. This shows that the proposed objective function is robust with respect to different observation noises.

Then, we illustrate the time series of 6 features to show how the simulated data resembles the synthetic data in Fig.~\ref{figure:simulationdata}. Specifically, we calibrate the model by $F_6$ and $f_x$ on the data generated based on the recommended parameter setting $\omega^*=[0.025,100,10,0.001,0.15,0.025]$ \cite{goosen2021calibrating}, where $f_x$ indicates any single feature which helps to calibrate the model and generate the best-simulated data. The simulated data calibrated by $F_6$ closely resembles the target data on all 6 features. To our best knowledge, this is the first ABM-agnostic method of successfully calibrating the high-frequency market time series at 1 second level and simulating 6 features with high-fidelity. In comparison, the simulated data of $f_x$ performs much poorer. This immediately suggests that calibrating more features can lead to low non-identifiability and high simulation fidelity.

\begin{table}[t]
\centering
\begin{threeparttable}
\caption{Linear regression coefficient table.\tnote{*}}
\label{table:differentepl}
\begin{tabular}{c||cccc}\hline\hline
{$\epsilon$}&{evaluation $D$} & slope & intercept & $R^2$\\ 
\midrule
\multirow{2}{*}{0.016} & W &-1.05  &9.44  &0.61 \\
\multirow{2}{*} &MSE &-0.185 &4.18 &0.59 \\
\midrule
\multirow{2}{*}{0.024} & W &-1.11 &9.28 &0.70\\
\multirow{2}{*} &MSE &-0.294 &4.66 &0.84  \\
\midrule
\multirow{2}{*}{0.028} & W &-0.746 &8.64 &0.54 \\
\multirow{2}{*} &MSE &-0.154 &4.18 &0.39 \\
\midrule
\multirow{2}{*}{0.030} &W &-0.570 &8.70 &0.61  \\
\multirow{2}{*} &MSE &-0.167 &4.48 &0.24  \\
\hline\hline
\end{tabular}
\begin{tablenotes}
\footnotesize
\item[*] Note that the unit of slope and intercept is at the $1E10^{-2}$ level.
\end{tablenotes}
\end{threeparttable}
\end{table}

\subsection{Ablation on Aggregation and Discrepancy}

The proof in Appendix~\ref{proof:uiqueness_of_max} shows that maximization uniquely preserves the intersection property of non-identifiable sets. To empirically validate this, Table~\ref{tab:aggregation_measure} presents an ablation study comparing aggregation methods (max, min, mean) and calibration measures (KL divergence (KL), Kolmogorov–Smirnov test (KS), MSE, W) across 10 synthetic datasets. 
The Friedman test confirms significant performance differences ($p = 2.88 \times 10^{-3}$). Subsequent Wilcoxon signed-rank tests show that max-W significantly outperforms min-W ($p = 4.88 \times 10^{-3}$) and mean-W ($p = 9.8 \times 10^{-4}$), empirically validating the theoretical claim that maximization enhances identifiability. Within the max group, max-W achieves the lowest rank (1.50), significantly outperforming KL ($p = 4.88 \times 10^{-3}$), KS ($p = 6.84 \times 10^{-3}$), and MSE ($p = 0.0527$), making W the preferable choice for FMS.

\begin{table}[t]
\centering
\begin{threeparttable} 
\caption{Statistical ablation tests of aggregation methods and calibration measures\tnote{*}.}
\label{tab:aggregation_measure}
\renewcommand{\arraystretch}{1.5}
\begin{tabular}{c|cccc|c|c}
\hline
aggregate   & \multicolumn{4}{c|}{max}                                  & min  & mean \\ \hline
measure & \multicolumn{1}{c|}{KL}  & \multicolumn{1}{c|}{KS}  
              & \multicolumn{1}{c|}{MSE} & W                             & W    & W    \\ \hline
data 1        & 3.19 & 4.47 & 3.47 & \textbf{3.07}                 & 6.63 & 4.77 \\ 
data 2        & 6.03 & 3.08 & 3.03 & \textbf{2.90}                 & 3.36 & 3.89 \\ 
data 3        & 6.86 & 4.30 & 3.19 & \textbf{3.08}                 & 4.63 & 3.47 \\ 
data 4        & 3.01 & 3.10 & 3.90 & \textbf{2.47}                 & 4.32 & 4.33 \\ 
data 5        & \textbf{2.55} & 6.58 & 3.75 & 2.84                  & 3.01 & 3.79 \\ 
data 6        & 3.71 & 6.69 & 4.76 & \textbf{3.67}                 & 3.74 & 5.73 \\ 
data 7        & 3.93 & 3.10 & \textbf{1.42} & 3.51                  & 3.23 & 3.52 \\ 
data 8        & 3.82 & 3.18 & 4.65 & \textbf{2.26}                 & 3.88 & 2.54 \\ 
data 9        & 8.35 & 5.48 & 3.85 & \textbf{3.51}                 & 5.07 & 5.40 \\ 
data 10       & 3.69 & 3.63 & \textbf{2.10} & 2.23                  & 5.32 & 4.04 \\ 
\#rank       & $3.90^+$ & $3.90^+$ & $2.90^{-}$ & \textbf{1.50}       & $4.30^{+}$ & $4.50^{+}$ \\ \hline
\end{tabular}
\begin{tablenotes}
\footnotesize
\item[*] Performance is at the $1E10^{-2}$ level. $+$ denotes significance under Bonferroni correction (family-wise $\alpha = 0.05$; $m = 5$, per-test $\alpha = 0.01$) in pairwise comparisons between max-W and others, and $^{-}$ indicates no significance.
\end{tablenotes}
\end{threeparttable} 
\end{table}

\begin{figure}[b]
  \centering
  \includegraphics[width=1\linewidth]{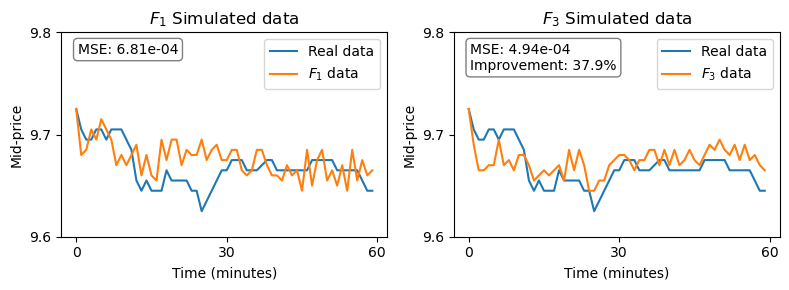}
  \caption{By calibrating with more features, the simulated data resembles the real market data better.}
  \label{figure:realdata}
\end{figure}

\subsection{Results on Calibration to the Real Data} 
At last, it is verified that the proposed method can also calibrate the real market data at a 3-second level. A 1-hour time series of 000001.sz is calibrated using $F_1$ and $F_3$. The simulated data generated by the corresponding best calibrated models is depicted in Fig.~\ref{figure:realdata}, together with the real observed data. Note that most of the FMS simulators (including the adopted PGPS model) model the order volume in a coarse-grained manner, i.e., simply setting all the orders with the identical volume of 100. Hence, only the mid-price time series is shown in a time step of 1 minute. It can be seen that the simulated mid-price of $F_3$ is more similar to the real observed data than $F_1$ by an improvement of 37.9\% in MSE. This confirms that the proposed method indeed can calibrate the real market data with multiple time series, thus suggesting a new effective calibration objective function for FMS. 

\section{Conclusions and Future Work}\label{section:conclusion}
This paper studies the non-identifiability issue of social simulation. To our best knowledge, this is the first work in which the non-identifiability is formally defined and theoretically alleviated. It is defined as the probability of random sampling in the non-identifiable set of the parameter space. It is rigorously analyzed that this probability can be reduced with more time series features in an exponential manner. Based on the analysis, a new objective function is proposed to effectively aggregate multiple features with a maximization function over existing discrepancy metrics. We theoretically and empirically show that the proposed aggregation is the optimal one. Extensive empirical studies have been conducted on 10 synthetic data and 1 real data of the FMS tasks. It has been successfully verified through simulations that, by using multiple features, not only can the non-identifiable set be minimized, but the simulation fidelity is significantly improved. Finally, we suggest how to select features using theoretical and empirical studies.

Future work can be multi-directional. First, while the study relies on handcrafted LOB features, learning-based methods, such as advanced representation learning techniques \cite{li2024simlob} and interpretable alpha factor discovery \cite{zhao2024quantfactor}, could help extract more meaningful market signals and refine feature selection. Second, developing adaptive recalibration methods that dynamically adjust parameters based on incoming market data could help maintain high-fidelity simulations over longer time horizons and under changing conditions. Third, beyond financial markets, the approach of reducing non-identifiability through multivariate feature calibration could extend to other agent-based simulations that generate multivariate time series data, like weather forecasting and traffic flow modeling.

\appendix

\section{Appendix}
\subsection{Proof of inequality~\eqref{eq:line2}}\label{appendix:proof1}
\begin{proof}\label{proof:2}
Recall that the inequality is
\begin{equation}\nonumber
\prod_{i=1}^{K}\beta_i \leq  \left(\frac{\sum_{i=1}^K \beta_i^2}{K}\right)^{\frac{K}{2}}.   
\end{equation}

$\forall i \in \{1,\cdots,K\}, $ we have $\beta_i >0$; by definition, $ \beta_i \leq 1$.

Consider $f(x) = \log_{e}^x \triangleq \ln x$, which is concave on $(0,+\infty)$. Suppose $p_i \geq 0$ and $\sum_{i=1}^Kp_i = 1$, by Jensen's inequality, we have
\begin{equation}
p_i \sum_{i=1}^K \ln x_i \leq \ln \left(\sum_{i=1}^K p_i x_i\right). 
\end{equation}

Replace $p_i$ and $x_i$ with $\frac{1}{K}$ and $\beta_i^2$, respectively, then
\begin{equation}\label{equation:proof2_Jensen}
\frac{1}{K}\sum_{i=1}^K \ln \beta_i^2 \leq \ln \left(\frac{1}{K}\sum_{i=1}^K \beta_i^2\right). 
\end{equation}

Also
\begin{equation}\label{equation:proof2}
\frac{1}{K}\sum_{i=1}^K \ln \beta_i^2 = \sum_{i=1}^K \ln \beta_i^{\frac{2}{K}} = \ln \left(\ \left(\prod_{i=1}^K \beta_i\right)^{\frac{2}{K}}\right).   
\end{equation}

By combining (\ref{equation:proof2_Jensen}) with (\ref{equation:proof2}), and noticing the fact that $\ln x$ is strictly increasing on $(0, +\infty)$, we get the desired result.\hfill $\blacksquare$
\end{proof}

\subsection{Proof of Theorem~\ref{def:expall}.}\label{appendix:proof2}
\begin{proof}\label{proof:3}
To prove the exponential alleviation of the upper bound of non-identifiability under the constraint
\begin{equation}\label{proof3:beta_constrain}
\beta_{K}^{2} < \frac{{\sum}_{i=1}^{K-1}\beta_i^2}{K-1},    
\end{equation}
we consider two cases with $K-1$ and $K$ features, respectively. The non-identifiability bounds in these cases are given by
 \begin{equation}\nonumber
P\left(\omega \in\{\mathcal{S}^{D,k,\epsilon}\}_{k=1}^{K-1} \middle| \hat{\mathbf{X}}_T\right) \leq \left(\frac{1}{K-1}\sum_{i=1}^{K-1} \beta_i^2\right)^\frac{K-1}{2}
\end{equation}
and
\begin{equation}\nonumber
P\left(\omega \in\{\mathcal{S}^{D,k,\epsilon}\}_{k=1}^K \middle| \hat{\mathbf{X}}_T\right) \leq \left(\frac{1}{K}\sum_{i=1}^{K} \beta_i^2\right)^{\frac{K}{2}}. 
\end{equation}

We then examine the ratio of these two bounds:
\begin{equation}\label{proof3:fraction}
\begin{aligned}
&\frac{\left(\frac{1}{K}\sum_{i=1}^{K} \beta_i^2\right)^{\frac{K}{2}}}{\left(\frac{1}{K-1}\sum_{i=1}^{K-1} \beta_i^2\right)^\frac{K-1}{2}} =\\
&\left(\frac{K-1}{K} \cdot \left(1+\frac{\beta_K^2}{\sum_{i=1}^{K-1} \beta_i^2}\right)\right)^{\frac{K-1}{2}} \cdot \left(\frac{\sum_{i=1}^{K} \beta_i^2}{K}\right)^{\frac{1}{2}}. \\
\end{aligned}
\end{equation}

Given that $\beta_i \in (0,1]$ for $i=1,\cdots, K-1$, we have
\begin{equation}
0 < \left(\frac{\sum_{i=1}^{K} \beta_i^2}{K}\right)^{\frac{1}{2}} \leq 1.  
\end{equation}

Furthermore, from (\ref{proof3:beta_constrain}), it follows that
\begin{equation}
0< \frac{K-1}{K} \cdot \frac{\beta_K^2}{\sum_{i=1}^{K} \beta_i^2} < \frac{1}{K}
\end{equation}
and therefore
\begin{equation}
0 < \frac{K-1}{K} \cdot \left(1+\frac{\beta_K^2}{\sum_{i=1}^{K-1} \beta_i^2}\right) < 1.
\end{equation}
  
Define
\begin{equation}\label{proof3:Delta_K}
\Delta_{K}=1-\frac{K-1}{K} \cdot \left(1+\frac{\beta_K^2}{\sum_{i=1}^{K-1} \beta_i^2}\right),  
\end{equation}
then $0< \Delta_{K} <1$. Substituting into the $RHS$ of (\ref{proof3:fraction}) yields
\begin{equation}\label{proof3:Delta_inequality}
LHS \ of \ \eqref{proof3:fraction} < \left(1-\Delta_{K}\right)^{\frac{K}{2}}.
\end{equation}

Let $\lambda_K \triangleq -\frac{1}{2} \ln (1 - \Delta_K) > 0$. Then $(1 - \Delta_K)^{\frac{K}{2}} = e^{-\lambda_K K}$. Since $\inf_K \Delta_K > 0$, it follows that $\inf_K \lambda_K > 0$. Define the \emph{uniform constant} $\lambda = \inf_K \lambda_K > 0$, so that $\lambda_K \geq \lambda$ for all $K$, and hence $e^{-\lambda_K K} \leq e^{-\lambda K}$. Multiplying both sides of~\eqref{proof3:Delta_inequality} by the LHS denominator in~\eqref{proof3:fraction} (which is less than $1$), and applying the above bound, we obtain
\begin{equation}
\left(\frac{1}{K}\sum_{i=1}^{K} \beta_i^2\right)^{\frac{K}{2}} < \left(1-\Delta_{K}\right)^{\frac{K}{2}} \leq e^{-\lambda \cdot K}.
\end{equation}

\hfill $\blacksquare$
\end{proof}

\subsection{Proof of Uniqueness of Max Aggregation}\label{proof:uiqueness_of_max}
\begin{proof}
We consider a common domain $\Omega^{D}$ for all $\Omega^{D,k}$, as originally stipulated in the definition~\ref{def:NIP_multile_dimension}.

Define $F_{\max}\{D_k\} = \max_{1 \leq k \leq K} D_k$, where 
\begin{equation}
D_k(\omega) \triangleq D(\hat{\mathbf{X}}_T^k, M(\omega))   
\end{equation} 
for clarity of notation. We first establish the utility of $F$, followed by its uniqueness. 

If $\omega \in \mathcal{S}^{F_{\max},\epsilon}$, then:
\begin{equation}
\begin{aligned}
&\max_{k} D_k(\omega) \leq \epsilon \implies D_k(\omega) \leq \epsilon,\ \forall k \implies\\
&\omega \in \bigcap_{k=1}^K \mathcal{S}^{D,k,\epsilon} \implies \mathcal{S}^{F_{\max}} \subseteq \bigcap_{k=1}^K \mathcal{S}^{D,k,\epsilon}
\end{aligned}
\end{equation}

If $\omega \in \bigcap_{k=1}^K \mathcal{S}^{D,k,\epsilon}$, then:
\begin{equation}
\begin{aligned}
&D_k(\omega) \leq \epsilon,\ \forall k \implies \max_{k} D_k(\omega) \leq \epsilon \implies \\
&\omega \in \mathcal{S}^{F_{\max},\epsilon} \implies \bigcap_{k=1}^K \mathcal{S}^{D,k,\epsilon} \subseteq \mathcal{S}^{F_{\max},\epsilon}
\end{aligned}
\end{equation}

Combine the results, we have
\begin{equation}\label{eq:set_equivalence}
\mathcal{S}^{F_{\max},\epsilon} = \bigcap_{k=1}^K \mathcal{S}^{D,k,\epsilon},\quad \forall \epsilon >0, \epsilon \in \mathbb{R}
\end{equation}
which completes the proof of the utility of $F$. Next, we establish the uniqueness of $F$ via contradiction. Suppose these exists $F' \neq F_{\max}$ that satisfies \eqref{eq:set_equivalence}. Consider two cases:

$\exists \{D_k^*\}$ s.t. $F'(\{D_k^*\}) < \max_k D_k^*$. Take $\epsilon_0 = F'(\{D_k^*\})$:
\begin{equation}
\begin{aligned}
&\max_k D_k^* > \epsilon_0 \ \text{but} \ \omega \in \mathcal{S}^{F',\epsilon_0} \\
&\omega \in \mathcal{S}^{F',\epsilon_0} \implies \omega \in \bigcap_{k=1}^K \mathcal{S}^{D,k,\epsilon_0} \implies D_k^* \leq \epsilon_0, \ \forall k .   
\end{aligned}
\end{equation}

Contradicts $\max_k D_k^* > \epsilon_0$.

$\exists \{D_k^*\}$ s.t. $F'(\{D_k^*\}) > \max_k D_k^*$. Take $\epsilon_0 = \max_k D_k^*$:
\begin{equation}
\omega \in \bigcap_{k=1}^K \mathcal{S}^{D,k,\epsilon_0} \implies D_k^* \leq \epsilon_0,\ \forall k \implies F'(\{D_k^*\}) \leq \epsilon_0.
\end{equation}

Contradicts $F'(\{D_k^*\}) > \epsilon_0$, thus completing the proof and establishing that maximization is the unique aggregation method that preserves the intersection of non-identifiable sets.

Moreover, we analyze two common alternative aggregation methods. The minimum aggregation function is defined as $F_{\min}\{D_k\} = \min_{1 \leq k \leq K} D_k$, This leads to
\begin{equation}
\mathcal{S}^{F_{\min}, \epsilon} = \bigcup_{k=1}^{K} \mathcal{S}^{D, k, \epsilon}.
\end{equation}

The weighted sum aggregation function is defined as $F_{sum} \{D_k\} = \sum_{k=1}^{K} \eta_k D_k$,
where $\eta_k > 0$ and $\sum_{k=1}^{K} \eta_k = 1$. This results in
\begin{equation}
\mathcal{S}^{F_{sum}, \epsilon} \subseteq \bigcap_{k=1}^{K} \mathcal{S}^{D, k, \epsilon / \eta_k}.
\end{equation}
\hfill $\blacksquare$
\end{proof}

\subsection{Proof of Selecting Linearly Independent Features}
\begin{proof}
We demonstrate that if two features are correlated by an affine transformation, there exists a distance metric for which their non-identifiable sets are identical. Therefore, features should be chosen to be linearly independent.

Let $\hat{\mathbf{X}}_T^k$ and $\mathbf{X}_T^k$ (for $k = 1, 2$) represent target and generated time series features related by:
\begin{equation}
\begin{cases} 
\hat{\mathbf{X}}_T^1 = a\hat{\mathbf{X}}_T^2 + b \\
\mathbf{X}_T^1 = a\mathbf{X}_T^2 + b 
\end{cases}, \quad a \neq 0, \ b \in \mathbb{R}.
\end{equation}

The non-identifiable sets are:
\begin{equation}
S^{D,k,\epsilon} = \left\{\omega \middle| D\left(\hat{\mathbf{X}}_T^k(\omega), \mathbf{X}_T^k\right) \leq \epsilon\right\}.
\end{equation}

Choose $D$ as the standardized MSE distance, denoted as $D_{\text{std}}(X,Y)$,
\begin{equation}
D_{\text{std}}(\mathbf{X}_T^k,\mathbf{Y}_T^k) \triangleq \frac{1}{T} \sum_{t=t_s}^{t_e} \left( \frac{X_t - \mu_{\mathbf{X}_T^k}}{\sigma_{\mathbf{X}_T^k}} - \frac{Y_t - \mu_{\mathbf{Y}_T^k}}{\sigma_{\mathbf{Y}_T^k}} \right)^2
\end{equation}
and the mean and standard deviation is defined as
\begin{equation}
\begin{aligned}
& \mu_{\mathbf{X}_T^k} \triangleq \frac{1}{T}\sum_{t=t_s}^{t_e} X_t \\
& \sigma_{\mathbf{X}_T^k} \triangleq \sqrt{\frac{1}{T}\sum_{t=t_s}^{t_e} (X_t - \mu_{\mathbf{X}_T^k})^2}.
\end{aligned}    
\end{equation}

For affine-related features, we have
\begin{equation}
\frac{\mathbf{X}_T^1 - \mu_{\mathbf{X}_T^1}}{\sigma_{\mathbf{X}_T^1}} = \text{sign}(a)\frac{\mathbf{X}_T^2 - \mu_{\mathbf{X}_T^2}}{\sigma_{\mathbf{X}_T^2}}. 
\end{equation}
the relationship for $\hat{\mathbf{X}}_T^2$ and $\mathbf{X}_T^2$ is analogous, which yields
\begin{equation}
D_{\text{std}}\left(\hat{\mathbf{X}}_T^1, \mathbf{X}_T^1\right) = D_{\text{std}}\left(\hat{\mathbf{X}}_T^2, \mathbf{X}_T^2\right).
\end{equation}

Consequently, the non-identifiable sets are identical, i.e.,
\begin{equation}
S^{D_{\text{std}},1,\epsilon} = S^{D_{\text{std}},2,\epsilon}. 
\end{equation}

Under the  max-aggregation $F$, we obtain 
\begin{equation}
S^{F,\epsilon} = S^{D_{\text{std}},1,\epsilon} \cap S^{D_{\text{std}},2,\epsilon} = S^{D_{\text{std}},k,\epsilon}, \quad \forall k.
\end{equation}

Thus, the measure remains unchanged:
\begin{equation}
\mu(S^{F,\epsilon}) = \mu(S^{D_{\text{std}},k,\epsilon}) \quad \forall k.
\end{equation}
\hfill $\blacksquare$
\end{proof}

\bibliographystyle{IEEEtran}
\bibliography{refs.bib}

\begin{IEEEbiography}[{\includegraphics[width=1in,height=1.25in, clip,keepaspectratio]{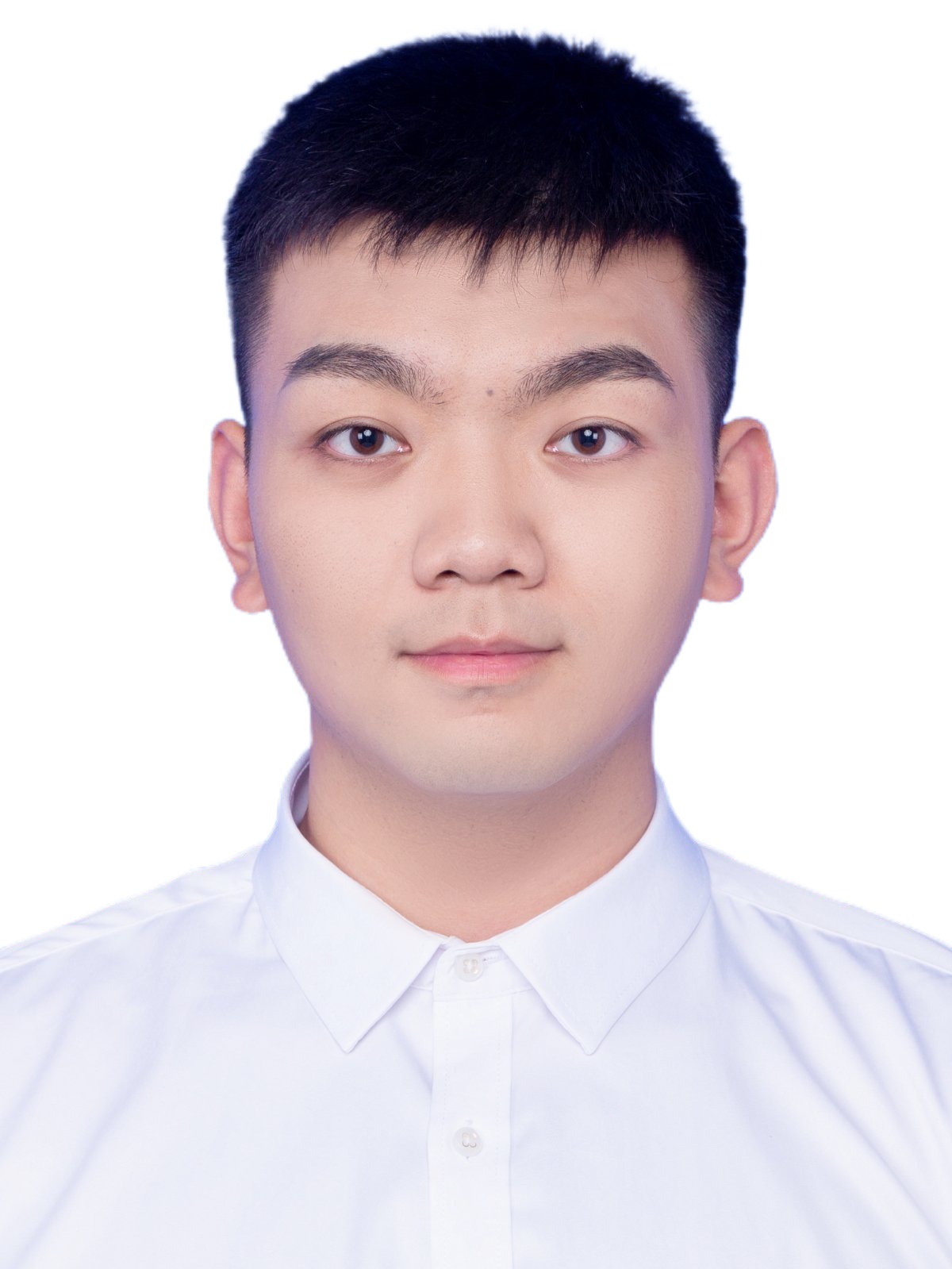}}]{Chenkai Wang} received his B.S. degree in Statistics and M.S. degree in Mathematics from Southern University of Science and Technology (SUSTech) in 2022 and 2024, respectively. His research interests include evolutionary computation and multi-agent systems.
\end{IEEEbiography}

\begin{IEEEbiography}[{\includegraphics[width=1in,height=1.25in, clip,keepaspectratio]{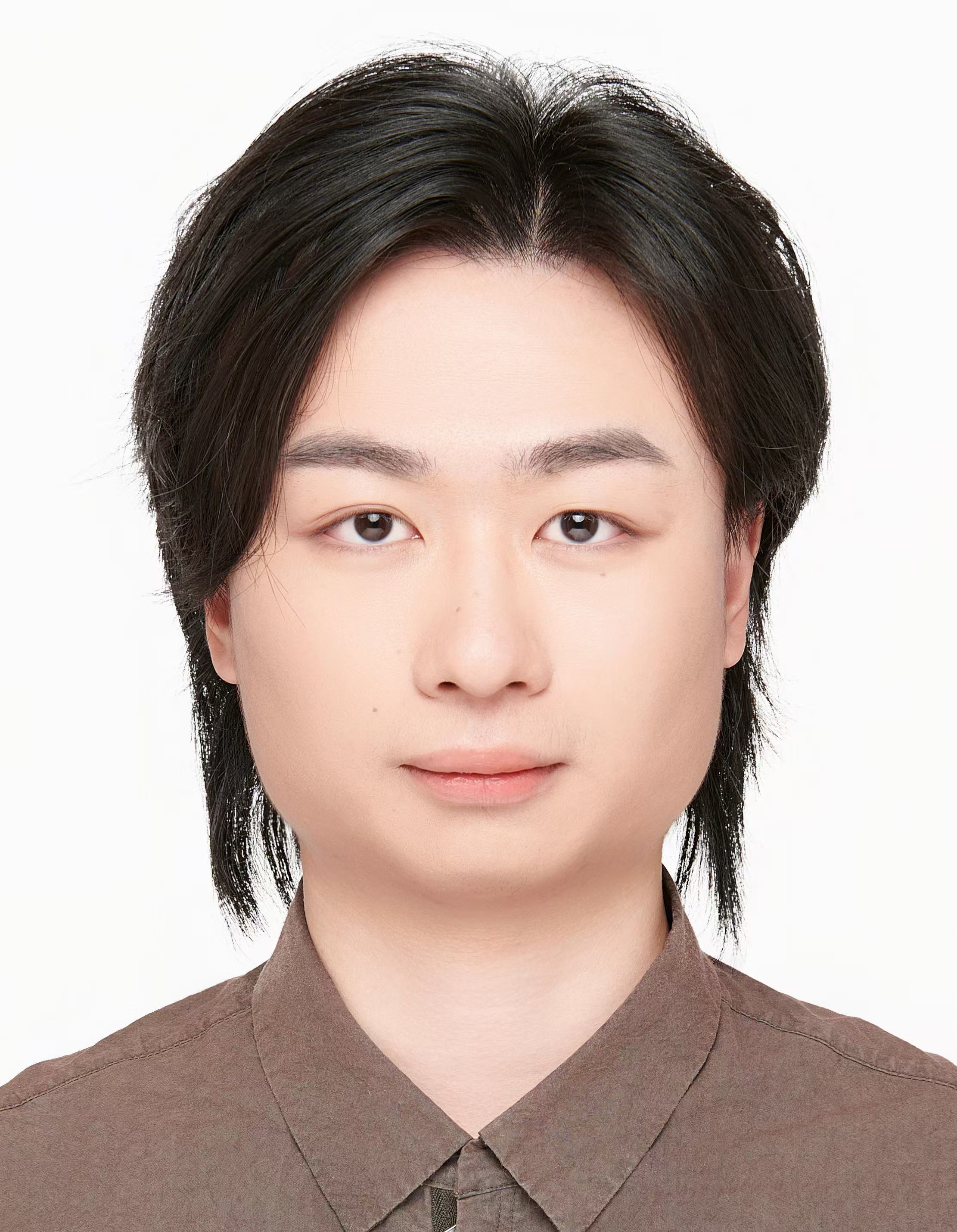}}]{Junji Ren}  received his B.E. degree in the Department of Computer Science and Technology from Southern University of Science and Technology (SUSTech) in 2024. He is currently pursuing his M.S. degree at the Southern University of Science and Technology. His research interests include evolutionary computation and multi-agent systems.
\end{IEEEbiography}

\begin{IEEEbiography}[{\includegraphics[width=1in,height=1.25in, clip,keepaspectratio]{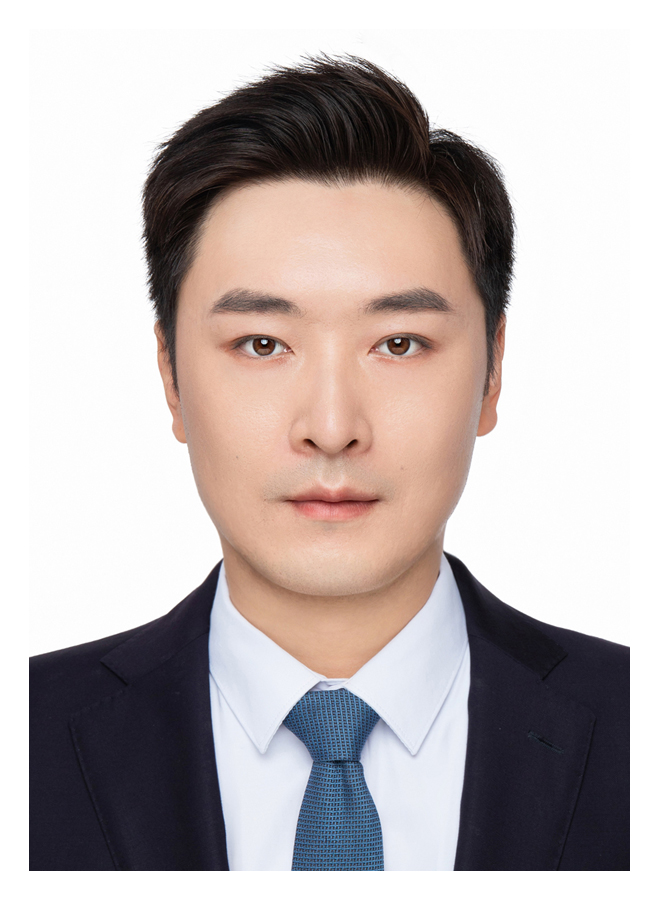}}]{Peng Yang} (S’14-M’18-SM’22) is a tenure-track assistant professor jointly in the Department of Statistics and Data Science and the Department of Computer Science and Engineering, Southern University of Science and Technology (SUSTech), China. He received his B.Sc. and Ph.D. degrees in the Department of Computer Science and Technology from the University of Science and Technology of China in 2012 and 2017, respectively. From 2017 to 2018, he was a Senior Engineer at Huawei, and then he joined SUSTech. His research interests include Reinforcement Learning, Quantitative Finance, and Multi-agent Systems. He has published over 40 papers in top journals and conferences like IEEE JSAC, IEEE TKADE, IEEE TEVC, IEEE TRO, IEEE TNNLS, IEEE TCYB, IEEE TCSS, and NeurIPS. He has served as the reviewer for top-tier journals (IEEE TEVC, TNNLS, TIE) and the PC member of top conferences (NeurIPS, ICLR, and ICML). He is the vice chair of the IEEE CIS Evolutionary Learning Task Force, an executive member of the CCF Computational Economics Committee and CCF AI Multi-agent System Committee, and a member of the IEEE CIS Evolutionary Computation Technical Committee. 
\end{IEEEbiography}

\end{document}